\documentclass[onecolumn]{IEEEtran}       % onecolumn (second format)
\usepackage{amssymb}
\usepackage{amsmath}
\usepackage{color}
\usepackage{tabularx}
\usepackage{lscape}
\usepackage{subfigure}
\usepackage{rotating}
\usepackage{amsthm}

\newtheorem{Theorem}{Theorem}
\newtheorem{Corollary}{Corollary}
\newtheorem{Definition}{Definition}

\newtheorem{Conjecture}{Conjecture}

\begin{document}

\title{Enhancing Performance Bounds of Multiple-Ring Networks with Cyclic Dependencies based on Network Calculus\footnote{A preliminary version of this paper appeared at RTCSA 2017 \cite{rtcsa-amari-17}}}

\author{\IEEEauthorblockN{Ahmed Amari } %1\textsuperscript{st} 
\IEEEauthorblockA{\textit{Complex Systems Engineering Dept} \\
\textit{ISAE -- Universit\'e de Toulouse}\\
Toulouse, France \\
ahmed.amari@isae.fr@isae.fr}
\and
\IEEEauthorblockN{Ahlem Mifdaoui} %2\textsuperscript{nd} 
\IEEEauthorblockA{\textit{Complex Systems Engineering Dept} \\
\textit{ISAE -- Universit\'e de Toulouse}\\
Toulouse, France \\
ahlem.mifdaoui@isae.fr}
}

\maketitle

\begin{abstract}
Tightening performance bounds of ring networks with cyclic dependencies is still an open problem in the literature. In this paper, we tackle such a challenging issue based on Network Calculus. First, we review the conventional timing approaches in the area and identify their main limitations, in terms of delay bounds pessimism. Afterwards, we have introduced a new concept called Pay Multiplexing Only at Convergence points (PMOC) to overcome such limitations. PMOC considers the flow serialization phenomena along the flow path, by paying the bursts of interfering flows only at the convergence points. The guaranteed end-to-end service curves under such a concept have been defined and proved for mono-ring and multiple-ring networks, as well as under Arbitrary and Fixed Priority multiplexing. A sensitivity analysis of the computed delay bounds for mono and multiple-ring networks is conducted with respect to various flow and network parameters, and their tightness is assessed in comparison with an achievable worst-case delay. A noticeable enhancement of the delay bounds, thus network resource efficiency and scalability, is highlighted under our proposal with reference to conventional approaches. Finally, the efficiency of the PMOC approach to provide timing guarantees is confirmed in the case of a realistic avionics application.
\end{abstract}
\begin{IEEEkeywords}
Network Calculus, ring, Cyclic dependencies, Serialization, Arbitrary Multiplexing, Fixed-Priority Multiplexing, Stability condition, Delay bounds

\end{IEEEkeywords}

%=========================================================================================================================================================================================
\section{Introduction}
\label{intro}
%what is the problem
During the last decades, ring-based networks have emerged in many real-time application domains, such as in automotive with the new standard TSN \cite{TSN}, automation with EtherCAT\cite{ethercat} and avionics with AeroRing \cite{amari2016aeroring}. Such a topology has interesting features to guarantee high availability level through the various redundancy solutions, specified in the documents IEC 624439/2-7 \cite{IEC62439-2} \cite{IEC62439-3} \cite{IEC62439-6} \cite{IEC62439-7}. The ring provides actually an implicit redundant path by introducing only one connection between the two end nodes, compared to line or star topologies \cite{Kleinberg-Etfa2010}. Moreover, ring topologies limit the cabling complexity; thus an inherent weight and installation costs reduction. A fundamental issue for such networks is bounding the timing performance to prove predictability, a key requirement for safety-critical applications. To deal with such a challenging need, an accurate timing analysis approach to compute worst-case delays or at least upper bounds is required.

%Why is it difficult
Many challenges arise from conducting such an analysis. First, while the implementation of event-triggered communication scheme on top of ring topologies offers high resource utilization efficiency and (re)configuration flexibility, it induces at the same time cyclic dependencies, i.e., there exist interfering flows with paths forming cycles. The impact of these cyclic dependencies on worst-case delays needs to be integrated. Second, the nodes connected via a ring topology generally implement service policies, such as First In First Out (FIFO), Fixed Priority (FP) and Arbitrary multiplexing, which inherently impact the worst-case delays. Third, there are diverse ring topologies varying from mono-ring to multiple-ring ones, which have to be considered during the analysis.

%What are the limitations of existent work
Among analytical methods to conduct worst-case performance analysis of ring networks, only few techniques have been proposed in the literature, mainly based on Network Calculus \cite{le2001network}. The high modularity and scalability of such a framework make it particularly efficient for complex communication networks \cite{Perathoner08}. It has been actually used for the certification of safety-critical networks, such as in avionics \cite{Grieu04,ahlem2010} and space \cite{Thomas11}. Existing approaches for ring networks are based on \textit{iterative local analysis}, by successively computing the delay bound in each crossed node either directly, i.e., \textit{Delay-based} methods \cite{cruz1991calculus}\cite{charny2000delay} \cite{Thiele08}, or from the backlog bound, i.e., \textit{Backlog-based} methods \cite{Tassiulas96} \cite{le2001network}; and summing these delays up results in end-to-end delay bounds. However, these approaches lead to overly pessimistic upper bounds, which decrease the network scalability and resource efficiency as it will be illustrated in Section \ref{classic-analysis}. \\

%Main contributions
Our main contributions are at both fundamental and practical levels:
\begin{itemize}
\item First, we review the conventional timing approaches in the area and identify their main limitations, in terms of delay bounds pessimism (for the Backlog-based approach \cite{le2001network}) or the limiting network stability condition (for the Time Stopping method \cite{cruz1991calculus});

\item Second, we tighten the delay bounds of ring networks with cyclic dependencies using Network Calculus, through introducing a new concept called Pay Multiplexing Only at Convergence points (PMOC). PMOC consists in considering the flow serialization phenomena along the path of a \textit{flow of interest (f.o.i)}, by paying the bursts of interfering flows only at the convergence points\footnote{Two flow paths may join at a node, called the convergence point, then disjoin after having a common subpath to maybe join again at another convergence point.}. The guaranteed end-to-end service curves for a f.o.i along its path under such a concept are defined and proved for mono-ring and multiple-ring networks under Arbitrary Multiplexing (Theorem \ref{Th:PMOO-Cycle} and Corollary \ref{th:GPMOC}) and FP multiplexing (Corollaries \ref{Th:PMOO-Cycle-FP} and \ref{cor:PMOC-FP}). Furthermore, the delay bounds computation is detailed and illustrated in a special case of ring networks, called \textit{regular ring networks}, for which the computation of end-to-end delay bounds can be considerably simplified, under specific necessary and sufficient conditions (Conjecture \ref{NSC-regular-def});

\item Third, we conduct a deep sensitivity analysis of the computed delay bounds for mono and multiple-ring networks with respect to various flow and network parameters, e.g., burst, rate, path length, network size and utilization rate. Moreover, we assess their tightness in comparison with an achievable worst-case delay (also called Worst-Case Delay (WCD) Lower Bound). We also benchmark the related work results against ours and highlight a noticeable enhancement of the delay bounds, thus network resource efficiency and scalability;

\item Finally, the efficiency of our proposal to provide timing guarantees is illustrated in the case of a realistic avionics application.\\
\end{itemize}

%paper organization
The rest of the paper is organized as follows: we start by presenting the main concepts of the Network Calculus framework in Section \ref{Background}, and detailing the main system assumptions and model in Section \ref{Model}. Then, we present the main iterative conventional Network Calculus approaches to compute the end-to-end delay bounds for ring networks with cyclic dependencies, and we show through a test case their limitations, in terms of network scalability (number of interconnected nodes) and resource efficiency (network utilization rate) in Section \ref{classic-analysis}. Afterwards, in Section \ref{Analysis}, we first introduce and prove our new timing analysis approach, PMOC, to enable the computation of tighter end-to-end delay bounds for ring networks. Extensive analyses of the proposed approach are conducted, regarding the delay bound tightness and its impact on the system performance, in comparison to conventional methods and a WCD lower bound. Section \ref{general-analysis} extends our proposed approach to the multiple-ring case and complements the conducted sensitivity analysis in Section \ref{Analysis}. Finally, we validate our proposal through a realistic avionics case study in Section \ref{UseCase}.

%===============================================================================================================
\section{Network Calculus Background}
\label{Background}

In this section, we present an overview of the main principles of Network Calculus \cite{le2001network} framework used in this paper. Further details on this framework can be found in two substantial books \cite{le2001network} and \cite{Chang2000}. The \textit{Network Calculus} is a mathematical framework to derive maximum bounds on system performance, such as delays, backlogs or throughput. This framework has been founded by the seminal work of Cruz in \cite{cruz1991calculusA,cruz1991calculus}, and then extended with min-plus Algebra operations in \cite{Chang2000} and \cite{le2001network}. The latter extension is based on the idea of modeling the communication nodes as in conventional system theory, with an input function, a transfer function and an output function, where addition and multiplication are replaced by minimum and addition, respectively. 

We answer herein some primordial questions when applying Network Calculus to conduct performance analysis of a realistic network. The first one concerns modeling the input traffic. The second is about modeling the node specifications, to integrate their impact on the system performance. Finally, we explain how to deal with a network of nodes to compute end-to-end performance guarantees.

\subsection{Traffic Model}
\label{TMH}

Network Calculus describes data flows by means of cumulative functions, defined as the number of transmitted bits during the time interval $[0, t]$. These functions are non negative and wide sense increasing:
\begin{displaymath}
\mathcal{F}=\{ f: \mathbb{R}^+ \rightarrow \mathbb{R}^+ \mid f(0)=0, \forall t \geq s: f(t) \geq f(s) \}
\end{displaymath}

Consider a system $S$ receiving input data flow with a Cumulative Arrival Function (CAF), $A(t)$, and putting out the same data flow with a Cumulative Departure Function (CDF), $D(t)$. Furthermore, $S$ fulfills the causality condition, i.e., $\forall t \in \mathbb{R}^+, A(t) \geq D(t)$. These functions allow computing the main performance metrics, defined as:
\begin{Definition}
\label{backlog-def}
The flow \textit{backlog} at time $t$ is:
\[ q(t) = A(t)-D(t) \]
\end{Definition}

%%%%%%%%%%%%%%%%%%%%%%%%%%%%%%%%%%%%%%%%
\begin{Definition}
\label{delay-def}
The flow \textit{virtual delay} at time $t$ is:
\[ d(t) = \inf\{ \tau \geq 0: A(t) \leq D(t+\tau)\} \]
\end{Definition}

The backlog $q(t)$ and virtual delay $d(t)$ are simply the vertical and horizontal distances between the CAF and the CDF at instant $t$, respectively. To compute upper bounds on the worst case delay and backlog, we need to introduce one of the most fundamental concepts in Network Calculus, the maximum arrival curve. This curve provides an upper bound on the number of events, e.g., bits or packets, observed during any interval of time. 
This concept allows modeling a large panel of event arrival patterns, such as periodic, sporadic, with or without jitter or burst. 
\begin{Definition}
\label{AC-def}
(Arrival Curve)\cite{le2001network} A function $\alpha$ is an arrival curve for a data flow with the CAF $A$, iff:
\[ \forall t,s \geq 0, s\leq t, A(t) - A(s) \leq \alpha(t-s) \]
\end{Definition}

The arrival pattern necessary to define the maximum arrival curve can be obtained from traffic traces if any, or application specification. The latter is more common for real-time communication networks. The network designer generally specifies a traffic contract for each application, enforced using a leaky-bucket shaper, which guarantees for the controlled traffic a maximum burst $\sigma$ and a maximum rate $\rho$, i.e., the traffic flow is $(\sigma,\rho)$-constrained. In this case, the arrival curve is a concave affine curve, defined as $\gamma_{\sigma, \rho} (t)= \sigma+ \rho.t$ for $t >0$.

\subsection{Node Model}
\label{NMH}
To conduct worst-case performance analysis, we need to put constraints on the input traffic through the maximum arrival curve notion. In return, we need to guarantee a minimum offered service within crossed nodes to cover the worst-case behavior and infer upper bounds on performance metrics, e.g., backlog and delay. This is done through the concept of minimum service curve, which has been defined for the first time in the seminal work \cite{Agrawal99} and more recently adapted in \cite{le2001network} as following.
\begin{Definition}
\label{SC-def}
(Simple Minimum Service Curve) The function $\beta$ is the simple service curve for a data flow with the CAF $A$ and the CDF $D$, iff:
%\begin{equation}
\[ \forall t \geq 0,D(t) \geq \inf_{s \leq t} (A(t)+ \beta(t-s)) \]
%\end{equation}
\end{Definition}
A very useful and common model of service curve is the rate-latency curve $\beta_{R,T}$, with $R$ the minimum guaranteed rate and $T$ the maximum latency before starting the service. This rate-latency function is defined as follows:
\[ \beta_{R,T}(t)= [R(t-T)]^+ \]
Where $\left[x\right]^+$ is the maximum between $x$ and $0$. This service curve is easy to define in the case of \textbf{\textit{one input/output node}} serving one or many traffic flows coming from the same source and going to the same destination. However, to handle more realistic scenario with a network of nodes, implementing aggregate scheduling, which multiplexes the crossing flows at the input and demultiplexes them at the output, we need to define the \textbf{\textit{left-over service curve}} guaranteed to each traffic flow within each crossed node, considering the impact of the other traffic flows in contention, to infer the offered guarantees for each flow. The computation of such a left-over service curve depends on the implemented scheduling policy within each crossed node, and the most common ones are Arbitrary Multiplexing, First In First Out (FIFO) and Fixed Priority (FP). It is worth noting that this derivation needs strict service curve property in the general case, except for FIFO and Constant bit rate nodes.
%A minimum strict service curve is defined as:
\begin{Definition}
\label{def:strict-service-curve}
(Strict service curve) The function $\beta $ is a strict service curve for a data flow with the CDF $D(t)$, if for any backlogged period\footnote{A backlogged period $]s,t]$ is an interval of time during which the backlog is non null, i.e., $A(s)=D(s)$ and $\forall u \in ]s,t]$, $A(u)-D(u) > 0$} $]s,t]$, $D(t) - D(s) \geq \beta(t-s)$.
\end{Definition}

The main results concerning the left-over service curves computation are as follows:
\begin{Theorem} (Left-over service curve - Arbitrary Multiplex)\cite{bouillard2009service}
\label{thm:residual-service-curve}
 let $f_1$ and $f_2$ be two flows crossing a server that offers a strict service curve $\beta$ such that $f_1$ is $\alpha_1$-constrained, then the left-over service curve offered to $f_2$ is:
\begin{displaymath}
\beta_2= (\beta - \alpha_1)_{\uparrow}
\end{displaymath}
where  $f_{\uparrow}(t) = \max\{0,\sup_{0 \leq s \leq t} f(s)\}$ 
\end{Theorem}
%%%%%%%%%%%%%%%%%%%%%%%%%%%%%%%%%%%%%%%%%%%%%%%%%%%%%%%%%%%
\begin{Corollary}(Left-over service curve - FP Multiplex)\cite{bouillard2011packetization}
\label{cor:residual-sc}
Consider a system with the strict service $\beta$ and $m$ flows crossing it, $f_1$,$f_2$,..,$f_m$. The maximum packet length of $f_i$ is $l_{i,max}$ and $f_i$ is $\alpha_i$-constrained. The flows are scheduled by the non-preemptive fixed priority (NP-FP) policy, where $priority~f_i \succ priority~f_j \Leftrightarrow i < j$. For each $i \in \{2,..,m \}$, the strict service curve of $f_i$ is given by:
%\begin{equation}
\[ (\beta - \sum_{j <i} \alpha_j - \max_{k \geq i} l_{k,max})_{\uparrow} \]
%\end{equation}
\end{Corollary}

\subsection{Performance Bounds}
\label{PAN}

Knowing the arrival and service curves, one may compute the upper bounds on performance metrics for a data flow. Before detailing the main theorems in this part, let us define the main algebraic operations in Network Calculus, i.e., convolution and deconvolution of two functions $f,g \in \mathcal{F}$ :
\begin{itemize}
\item min-plus convolution: 
%\begin{equation}
\[ f\otimes g(t) = \inf_{0 \leq s \leq t} \{ f(s) + g(t-s)\}\]
%\end{equation}
\item min-plus deconvolution:
%\begin{equation}
\[f \oslash g(t) = \sup_{\forall u \geq 0} \{ f(t+u) - g(u)\} \]
%\end{equation}
\end{itemize}

For a node with \textbf{\textit{one input/output}}, these bounds are computed according to the following theorem. 
\begin{Theorem} (Performance Bounds)
\label{TH1}
Consider a flow constrained by an arrival curve $\alpha$ crossing a system $\mathcal{S}$ that offers a service curve $\beta$. The performance bounds obtained at any time $t$ are given by:\\
Output arrival curve: $\alpha^*(t) =\alpha \oslash \beta (t)$\\
Backlog\footnote{$v(f,g)$: the maximum vertical distance between $f$ and $g$}: $ \forall~t:~q(t)\leq (\alpha\oslash\beta)(0)=:v(\alpha,\beta)$ \\
Delay\footnote{$h(f,g)$: the maximum horizontal distance between $f$ and $g$}: $ \forall~t:~d(t)\leq \inf\{t\geq 0 : (\alpha\oslash\beta)(-t)\leq0\}=:h(\alpha,\beta)$ 
\end{Theorem}
The calculus of these bounds is greatly simplified in the case of a leaky-bucket arrival curve ($\gamma_{b,r}$) and a rate-latency service curve ($\beta_{R,T}$). In this case, the delay and backlog are bounded by $\frac{b}{R} + T$ and $ b + r*T$, respectively; and the output arrival curve is $b+r (T + t)$. 

Afterwards, to extend this result to a \textbf{\textit{network of nodes}}, one of the strongest result in the Network Calculus framework is the computation of an end-to-end service curve for a tandem of nodes crossed by the same flows. This curve is computed as the convolution of residual service curves in each node, and is used to infer end-to-end performance bounds according to Th. \ref{TH1}. This result is described in the following theorem.
\begin{Theorem} (Concatenation-Pay Bursts Only Once)
\label{thm:concat}
Assume a flow crossing two servers with respective service curves  $\beta_1$ and $\beta_2$. The system composed of the concatenation of the two servers offers a minimum service curve $\beta_1 \otimes \beta_2$ to the flow.
\end{Theorem}

As an example, for a tandem of nodes with rate-latency service curves, the end-to-end service curve computed according to Th. \ref{thm:concat} is also a rate-latency curve, where the rate is the minimum of the crossed node rates and the latency is the sum of their latencies.

This result infer an interesting property known as "Pay bursts Only Once Phenomena". Indeed, the end-to-end delay bound for a data flow, computed using the end-to-end service curve obtained with Th. \ref{thm:concat}, clearly outperforms the sum of delay bound per node, computed iteratively using Th. \ref{TH1} and denoted as additive delay bound. The computation of these two bounds show the appearance of the burst term many times in the additive delay bound, as opposed to only once for the other. More recently, the authors in \cite{Schmitt08} propose an innovative approach, denoted as Pay Multiplexing Only Once (PMOO), and the main idea is based on taking into account the flow serialization phenomena along the flow path to compute tighter end-to-end delay bound. However, the latter has been proved only under arbitrary multiplexing.

%================================================================================================================
\section{System Model}
\label{Model}

We are interested in computing an upper bound on Worst-Case Delay for a flow of interest (\textit{f.o.i}) in multiple-ring networks with cyclic dependencies. To conduct such a timing analysis, we first consider the case of a mono-ring network, afterwards we generalize our approach to multiple-ring networks. Hence, we present herein the considered assumptions and notations for the mono-ring network, which will be extended to multiple-ring networks in Sec. \ref{Model-ext}. We use upper indices to indicate nodes or a set of nodes, and lower indices to indicate flows.
%%==========================================================
\begin{figure}[h!]
	\centering
		\includegraphics[width=0.8\textwidth]{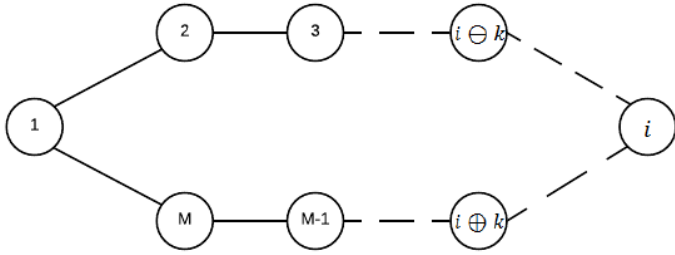}
	\caption{Ring network Example}
	\label{fig:generalNetwork}
\end{figure}
%%====================

\begin{itemize}

\item We consider a unidirectional ring topology, as shown in Fig. \ref{fig:generalNetwork}, connecting $M$ nodes, labelled from $1$ to $M$, and serving a fixed set of flows $I$. The unidirectional topology is not restrictive, since a full-duplex ring can be considered as two independent unidirectional rings that can be analyzed separately;

\item Each flow $i \in I$ follows a fixed path from its initial source until the final sink, defined as $\mathbb{P}_i= (0, i.ft, i.ft\oplus 1, ..., i.ft\oplus(h_i-1))$, where $0$ is a virtual node representing the source, $i.ft$ the first hop and $h_i$ the number of hops of flow $i$ with $h_i \leq M$ and the notations $l \oplus k$ and $l \ominus k$ designate the $k-th$ node downstream and upstream from node $l$, respectively, where the first downstream node for node $M$ is node $1$ and the first upstream node for node $1$ is node $M$. For a flow $i$, the specific case $i.ft\ominus1$ is the virtual node $0$. Moreover, we define its subpath through $n \in [1, h_i ]$ hops as $\mathbb{P}_i (n) = (0, i.ft, ..., i.ft\oplus(n-1))$, i.e., $\mathbb{P}_i=\mathbb{P}_i(h_i)$. It is worth noting that we consider only the output port of crossed nodes within the subpath $\mathbb{P}_i (n)$. Moreover, we assume that no two flows have the same path, since we can aggregate such flows (if any) and thus consider the aggregate flow;

\item Within the network, flows are treated according to an aggregate scheduling, i.e., flows are classified within aggregates according to a common parameter, such as priority. Within an aggregate, flows are served under arbitrary multiplexing in each crossed node;

\item We denote $i \ni k$ the set of flows crossing the node $k$, i.e., $i \ni k = \{ i \in I \mid k \in \mathbb{P}_i\}$; 

\item Consider $\mathbb{K}_f(n)$ the set of interfering flows with a \textit{f.o.i}. $f$ along its subpath $\mathbb{P}_f (n)$; so that $\mathbb{K}_f(n) = \{i \neq f / \exists k \in \mathbb{P}_f(n) / i \ni k \}$. Moreover, for any flow $i \in \mathbb{K}_f(n)$, consider its first (resp. last) multiplexing node label with flow $f$ along the subpath $\mathbb{P}_f (n)$ as $Mft(i, f, n)$ (resp. $Mlt (i,f, n)$);  

%\item Consider $conv(i,f,n)$ the convergence points of the $f.o.i$ $f$ with the interfering flow $i$ along its subpath of length $n$. In a network, two flows paths may join at a node, called a convergence point, then disjoin after having a common subpath to maybe join again at another convergence point and disjoin several times.

\item Each flow $i \in I$ has the CAF $A_i^k$ and the CDF $D_i^k$ at the node $k$;

\item Each flow $i \in I$ is constrained by one leaky bucket of rate $\rho_i$ and an initial burst $\sigma_i^0$ at its input source $0$, thus admits an initial input arrival curve $\alpha_i^0(t) = \sigma_i^0 + \rho_i.t$. Moreover, we define its input arrival curve at each crossed node $k$ along its path $\mathbb{P}_i$, as $\alpha_i^{k\ominus 1} (t) = \sigma_i^{k\ominus 1}+ \rho_i t$;

\item Each node $k$ serves the traffic of an aggregate according to a strict service curve having a rate-latency form, with a rate $R^k$ and a latency $T^k$, $\beta^k(t)=[R^k(t-T^k)]^+$;

\item We consider the case of networks where the following stability condition is satisfied: for any node $k \in [1, M ] $, $\frac{\sum_{i \ni k} \rho_i}{R^k} \leq 1$. This condition is necessary to guarantee finite delay bounds within each crossed node.
\end{itemize}
All notations are summarized in Tab. \ref{tab1} in appendices.

%###################################%###################################
\section{Conventional Analysis Methods and Limitations}
\label{classic-analysis}
%###################################%###################################

One of the major challenges in applying Network Calculus is improving accuracy of performance bounds to avoid over-dimensioning of network resources; thus increasing the integration costs. In the research community, there has been a growing interest in the subject and several approaches have been proposed to deal with the delay bounds tightness in networks with acyclic graph, also known as feedforward networks. An interesting overview of the most relevant approaches in this area is detailed in \cite{Fidler10}. However, only few approaches related to computing end-to-end delay bounds in non-feedforward networks, i.e., network with cyclic dependencies, are reported in the literature, and none of these are dealing with the tightness issue.

A first class of interesting approaches has been proposed to break the potential cycles through prohibiting the use of some links or sub-paths to ensure the feed-forward property \cite{schroeder91} \cite{Starobinsk03}. Although these approaches simplify the timing analysis of non-feedforward networks, they imply at the same time a reliability level deterioration, since the use of some links is forbidden, e.g., a ring topology is transformed into line.

The second class of approaches introduces computation methods to support cycles using an iterative approach by successively analyzing the delay bound in each crossed node in the network, resulting in end-to-end delay bounds computation. The most relevant approaches are focusing on, either each crossed node delay bound, e.g., \cite{cruz1991calculus} \cite{charny2000delay} \cite{Thiele08}, or each crossed node backlog bound, e.g., \cite{Tassiulas96} \cite{le2001network}. For the particular case of ring-based network, two interesting approaches have been proposed: the  \textbf{\textit{Time Stopping Method}} \cite{cruz1991calculus} and the \textbf{\textit{Backlog-based Method}} \cite{le2001network}.

In this section, we detail these two main conventional iterative analyses of delay bounds, based on Network Calculus. Then, we point out the limitations of each approach through an illustrative example.

\subsection{Time Stopping Method}
%###################################
\label{section:cruz}
This approach has been proposed in \cite{cruz1991calculus} and consists of two steps. First, a finite burstiness bound for transmitted flows is assumed to obtain a set of equations to compute the delay bounds. Then, the feasibility conditions to solve these equations are defined. Therefore, we will first express all the equations to compute the upper bounds on bursts and delays in each crossed node. Then, we deduce the feasibility condition.

In \cite{cruz1991calculus}, the burst propagation formula of a flow $i$ at the output of node $j$ is given by:
\[\sigma_{i}^j= \sigma_{i}^{j\ominus 1}+\rho_{i}*Delay^j\]
where $Delay^j$ is the delay within node $j$.

Hence, at the output of node $j$, flow $i$ has already crossed $(j-i)\mod M$ nodes since node $i$. The output burst of flow $i$ at the node $j$ is given as follows:
\begin{equation}
\sigma_{i}^j=\sigma_{i}^0+\rho_{i}*\sum_{k=0}^{(j-i)\mod M} Delay^{i\oplus k} 
\label{formula:Cruzburst}
\end{equation}

On the other hand, the delay $Delay^{k}$ within node $k$ to process the crossing traffic is equal to the sum of its latency $T^{k}$ and the processing time of all the crossing bursts:
\begin{equation}
Delay^{k}=\frac{\sum_{j\ni k} \sigma_{j}^{k\ominus 1}}{R^{k}} + T^{k}
\label{formula:Cruzdelay}
\end{equation}

Equations (\ref{formula:Cruzburst}) and (\ref{formula:Cruzdelay}) can be represented by the following matrix system:
\begin{equation}
\left\{ \begin{array}{c}
	D = A_1* B + C_1\\
	B = A_2*D + C_2
\end{array} 
\right.
\label{formula:CruzMatrix}
\end{equation}
where $D$ is the vector of delays, $B$ is the vector of propagated bursts, and $C1$ and $C2$ are the  constant vectors.

Thus, by propagating these constraints, we obtain:
\begin{equation}
D=[I-A_1*A_2]^{-1}*C_3
\label{formula:delayMatrix}
\end{equation}
where $C_3=A_1*C_2+C_1$ and $I$ is the identity matrix.

The system admits a solution if the $[I-A_1*A_2]$ matrix is invertible, i.e., its determinant is not null. If this condition is verified, the upper bounds on delays can be computed.

The end-to-end delay communication bound of a given flow $i$ along its path $\mathbb{P}_{i}$, $EED_i$, is defined as follows:
\begin{equation}
	EED_{i}=\sum_{k \in \mathbb{P}_{i}}{(Delay^{k}+\delta)}
	\label{formula:delayGeneral}
\end{equation}
where $\delta$ is the propagation delay. 	

%###################################
\subsection{Backlog-based Method}
%###################################

This method has been initially proposed in \cite{Tassiulas96} and more recently generalized in \cite{le2001network}. The authors provide the maximum backlog bound when considering non work-conserving nodes, which is a maximum bound on the total amount of data present in the network at any time. This maximum backlog bound within node $k$ is as follows:
\begin{equation}
Backlog^k=M\frac{\mu}{\eta}(M\sigma^{\max}+B)+\sigma+B
\label{formula:backlog}
\end{equation}

where:
\begin{itemize}
	\item $\sigma=\sum_i{\sigma_{i}^0}$ is the sum of all flows bursts, and $\sigma^{\max}=\max_{k} \sum_{j \ni i} \sigma_{j}^{k\ominus 1}$ is the maximal sum of bursts that pass through any node;
	\item $\mu=\max_{i} [\sum_{j \ni i} \rho_{j}] $;
	\item $\eta=\min_i(R^{i} - \sum_{j \ni i} \rho_{j})$;
	\item $B=\sum_i{R^{i}.T^{i}}$
\end{itemize}

The maximum bound on the delay within each node $i$ is the processing time of the maximum backlogged traffic $Backlog^i$ in Eq. (\ref{formula:backlog}) served with a transmission capacity $R^{i}$, and it is as follows:
\begin{equation}
	Delay^{i}=\frac{Backlog^i}{R^{i}}
	\label{formula:delaiGeneral}
\end{equation}

The end-to-end delay communication bound still is computed using Eq. (\ref{formula:delayGeneral}).

\subsection{Discussion}
\label{RW-results}
In this section, we detail some numerical results of the delay upper bounds of a single ring network, similar to the one illustrated in Fig. \ref{fig:generalNetwork}, based on both conventional analysis methods to point out their limitations. We consider the case study with the following assumptions:

\begin{itemize}
	\item The topology is a unidirectional ring topology, connecting $M$ nodes;
	\item All nodes are similar and each node has a service curve $\beta_{R=1Gbit/s,T=600ns}$;
	\item Each node generates a broadcast traffic ($\sigma=128 bytes, \rho=128 Kbps$)-constrained with a deadline of $1ms$.
\end{itemize}

Scenarios are generated varying the flow and network parameters, as follows:
\begin{itemize}
\item Network size is varying from $10$ to $100$ nodes with a step of $10$ nodes, i.e., $M \in [10, 100]$;
\item The maximum utilization rate is varying where $U_{max} \in [10\%, 100\%[$ with a step of $10\%$, through varying the flow rate according to the following condition: $\frac{M. \rho_{\max}}{R} \leq U_{max}$.
\end{itemize}

\begin{figure*}[htbp]
		\centering
		\includegraphics[scale=0.45]{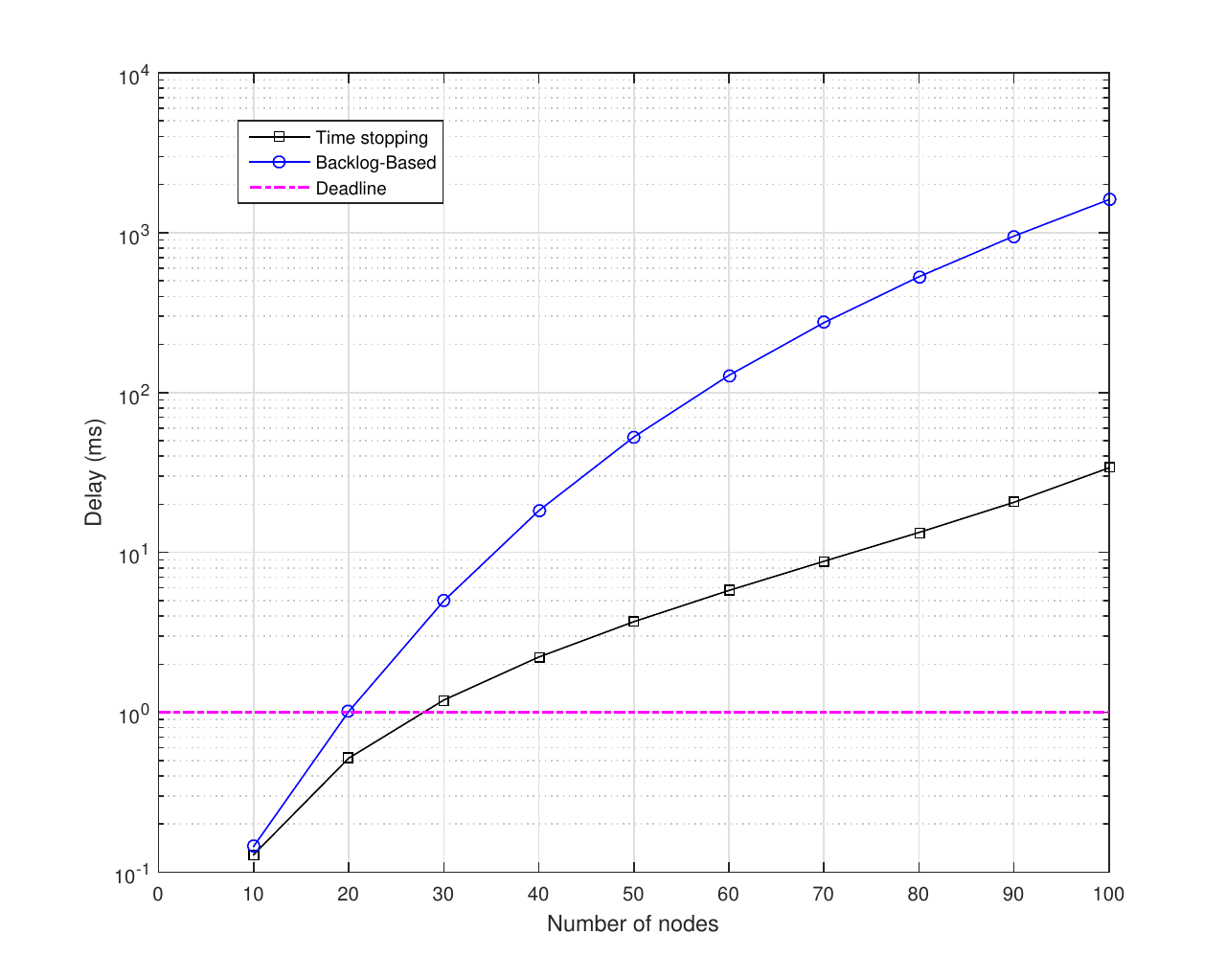}
	\caption{End-to-end delay bounds vs number of nodes.}
	\label{fig:RW-comparisonb}
\end{figure*}

Fig. \ref{fig:RW-comparisonb} shows a comparison of both approaches in terms of end-to-end delay bounds, when enlarging the network size. Obviously, the delay bounds increase with the network size, since the number of transmitted messages and crossed nodes increases. As we can notice, for a large-scale network, e.g., 100 nodes, both approaches do not respect the flows deadline ($1$ms) and guarantee pessimistic delay bounds, e.g., $33.8$ms and $1.6$s for Time Stopping and Backlog-based methods, respectively. Hence, the maximum network size respecting the flow deadline is about 20 and 27 nodes for the Backlog-based and Time Stopping methods, respectively.
%=================================================================================================
%===============================================================================
\begin{figure*}[htbp]
		\centering
		\includegraphics[scale=0.45]{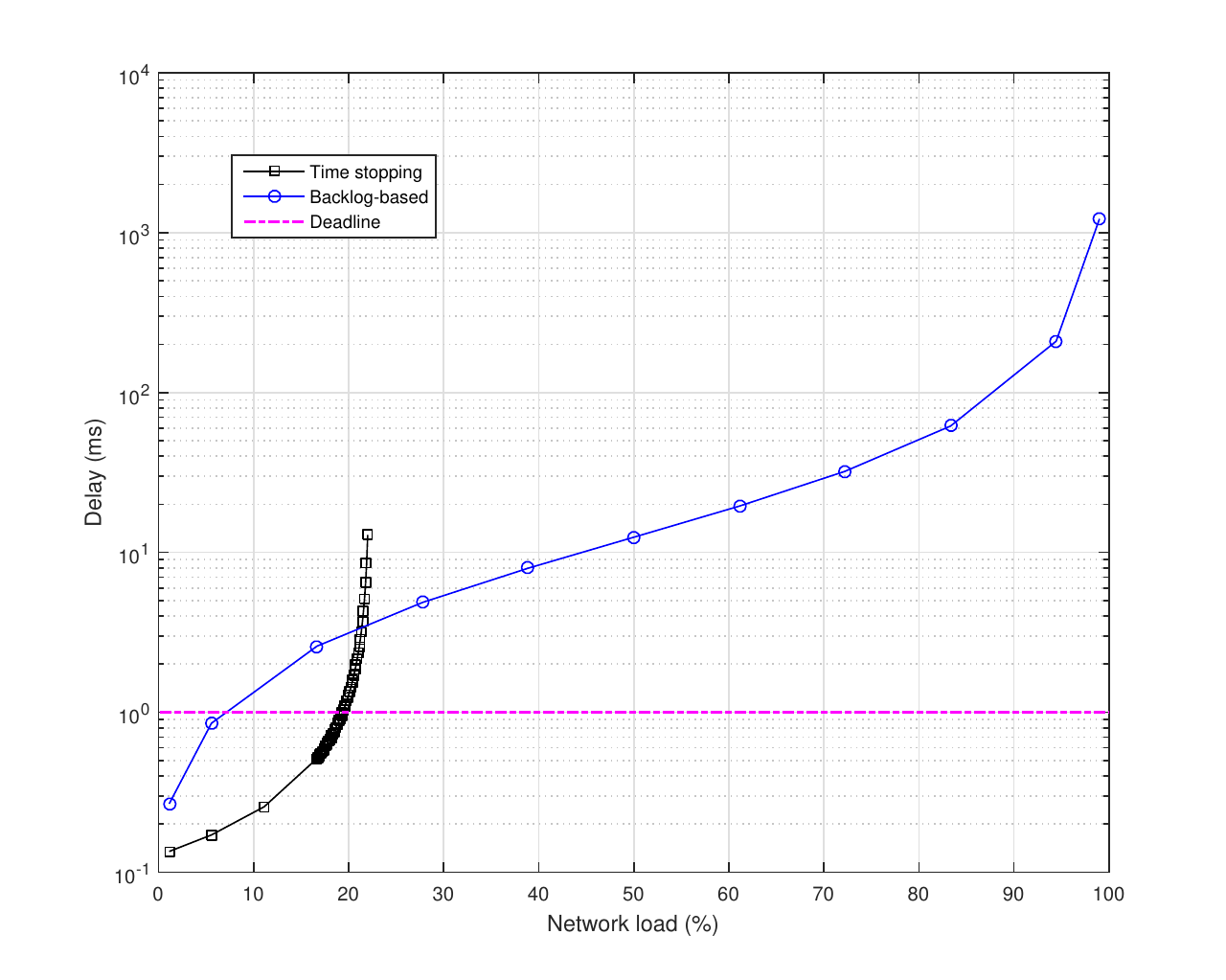}
	\caption{End-to-end delay bounds vs network utilization rate.}
	\label{fig:RW-comparisonc}
\end{figure*}
%==============================================================================

Fig. \ref{fig:RW-comparisonc} illustrates the impact of increasing the congestion on the end-to-end delay bounds under both methods. Obviously, the delay bound increases with the network load under both methods since the amount of transmitted data increases, which increases the interferences. As we can see, the Time Stopping method offers tighter bounds until it reaches its limit, i.e., it diverges for $U_{max}=22.22\%$; whereas the Backlog-based can achieve a full utilization rate, even if the delay bounds are overly pessimistic, e.g., $1,22$s for $U_{max}=99\%$. Moreover, the maximum network utilization rate respecting the flows deadline is only about $7.1\%$ and $19.36\%$ with the Backlog-based and Time Stopping methods, respectively.

These results have the following theoretical explanations. For Time Stopping method, the matrix $[I-A_1*A_2]$ is as follows:
%%%%%%%%%%%%%%%%%%%%%%%%%%%%%
\begin{equation}
-\frac{1}{R}\times\left(
\begin{array}{ccccc}
-R&\rho&2\rho&\cdots&M\rho \\
M\rho&-R&\rho&\cdots&(M-1)\rho \\
(M-1)\rho&M\rho&-R&\cdots&(M-2)\rho \\
\vdots&\vdots&\vdots&\ddots&\vdots \\
\rho&2\rho&3\rho&\cdots&-R \\
\end{array}\right)
\label{matrix:determ}
\end{equation}
%%%%%%%%%%%%%%%%%%%%%%%%%%%%%

\begin{figure*}[htbp]
		\centering
		\includegraphics[scale=0.65]{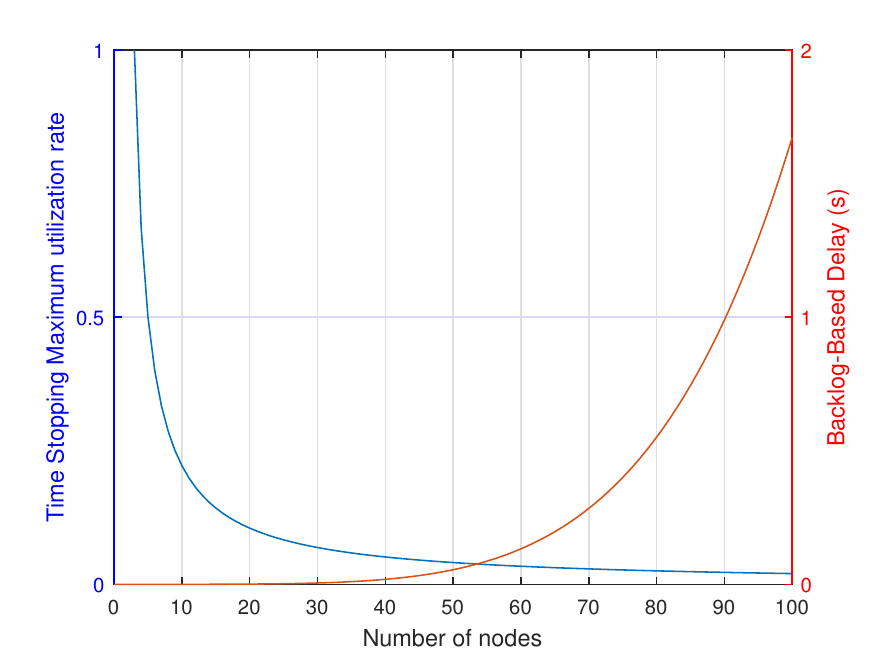}
	\caption{The maximum utilization rate for the Time Stopping Method and upper bound on delays for Backlog-based Method vs number of nodes.}
	\label{fig:RW-comparisona}
\end{figure*}

The system admits a solution if the matrix determinant is not null. In this particular case, the feasibility condition is $\rho < \frac{2*R}{M(M-1)}$. Therefore, the method allows computing bounds when the maximum utilization rate of the network is less than $\frac{2}{(M-1)}$. As we can see in Fig. \ref{fig:RW-comparisona}, the maximum utilization rate for the Time Stopping method tends to $0$, when $M \to \infty$, e.g., less than $0.1$ for $20$ nodes. This implies that the network has to be under utilized to satisfy the network stability condition, which limits the network resource-efficiency.

On the other hand, with the Backlog-based approach the backlog within any crossed node $k$ and the end-to-end delay bounds for any flow $i$ become polynomial functions of the variable $M$ (number of nodes) of degree 3, and 4, respectively: 
\[ Backlog^k=M\frac{\tau}{1-\tau}\cdot(M^2\times\sigma+M\times L)+M(\sigma+ L)\]
\[ EDD_i=M(\frac{Backlog^k}{R}+\delta) \]
where $\tau=\frac{M\times \rho}{R}$.

This fact implies an end-to-end delay bound growing as $\theta(M^4)$, as shown in Fig \ref{fig:RW-comparisona}. Hence, as we can notice, the Time Stopping approach offers tighter delay bounds than the Backlog-based approach when the network is stable, i.e., $U_{max}<\frac{2}{M-1}$. However, the Backlog-based approach can guarantee a full utilization rate, even if the delay increases dramatically.

\textbf{\textit{The Time Stopping method actually limits the network performance in terms of resource efficiency, i.e., the utilization rate decreases dramatically when the network size increases; whereas the Backlog-based method limits the system scalability, i.e., the nodes number is hardly constrained to guarantee the temporal deadlines.}}

To overcome these limitations, we introduce in the next section an enhanced worst-case timing analysis of ring-based networks with cyclic dependencies, taking into account the flow serialization phenomena along the flows paths.

%###################################%###################################
\section{Pay Multiplexing Only at Convergence Points}
\label{Analysis}
%###################################%###################################

This approach consists in considering the flow serialization phenomena along the path of a \textit{f.o.i}, by paying the bursts of interfering flows only at the convergence points. Similar concepts have been developed in the literature for feedforward networks, i.e., with no cyclic dependencies, such as the Pay Bursts Only Once (PBOO) in \cite{le2001network} and the Pay Multiplexing Only Once (PMOO) in \cite{Schmitt08} \cite{bouillard2008optimal}. However, tightening the delay bounds of non-feedforward networks still is an open problem in the literature, and such an approach does not exist yet for non-feedforward networks. The main idea of this method is to handle such an issue for ring-based networks. 

In the rest of this section, we detail the main idea and steps of the PMOC approach, to compute delay upper bounds in ring-based networks with cyclic dependencies. We thus introduce the main concept progressively through an illustrative example to highlight the cycle dependency problem. Then, we define and prove the closed-form service curve in a mono-ring network under Arbitrary and Fixed Priority multiplexing. Afterwards, based on these defined service curves, the necessary and sufficient condition to infer the computation of end-to-end delay bounds is defined in the general case and specified for a special case of ring networks. Finally, we conduct a performance evaluation of our approach under different scenarios to assess its sensitivity and tightness.

%###################################
\subsection{Illustrative Example}
%###################################
\label{sec:PMOCbasics}

%=======================================================================================
\begin{figure}[htbp]
\centering     %%% not \center
\includegraphics[width=75mm]{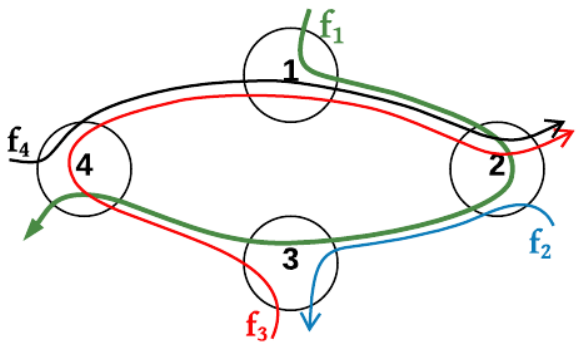}
\caption{A Ring network with cyclic dependency.}
\label{figure:cycle}
\end{figure}
%=====================================

We illustrate herein the cyclic dependency problem and the main idea of PMOC principle through the example of Fig. \ref{figure:cycle}. 

Consider as a \textit{f.o.i} $f_1$ with the path $\mathbb{P}_{f_1} = (0, 1, 2, 3)$. To compute the end-to-end delay bound of $f_1$, we need to integrate the impact of all the interfering flows along its path, $\mathbb{K}_{f_1}(3) = \{f_2, f_3, f_4 \}$. Hence, at the input of node $1$, we need to quantify the arriving bursts of flows $f_3$ and $f_4$. Moreover, the burst of $f_4$ at the input of node 1 depends on the burst of $f_3$ at the input of node 4, which in its turn depends on the burst of the \textit{f.o.i} $f_1$ at the input of node 3. As we can notice, to analyse the impact of interfering flows on the \textit{f.o.i} $f_1$, we need to quantify its impact on these interfering flows; thus the cyclic dependency. There is actually no start point, where all the flows bursts are known, to launch the delay computation. 

To overcome such a difficulty, the main idea of PMOC approach is to compute the tightest possible upper bound on these unknown bursts, when considering the flow serialization phenomena along the path of the \textit{f.o.i} and integrating the impact of interfering flows only at the convergence points. As illustrated in Fig. \ref{figure:cycle}, because of the ring topology, there are only two possible convergence points with a \textit{f.o.i}: 
\begin{itemize}
\item If the convergence point is the interfering flow source, then the burst impacting the \textit{f.o.i} is known, e.g., $f_2$ burst in node 2;
\item If the convergence point is the source of the \textit{f.o.i}, then the burst impacting the \textit{f.o.i} is unknown, e.g., $f_3$ and $f_4$ bursts in node 1. 
\end{itemize}

Consider the example of computing the unknown burst of $f_4$ at the input of node 1. To compute such a propagated burst, we need to quantify the minimum guaranteed service of $f_4$ until reaching the input of its convergence point with the \textit{f.o.i} $f_1$, i.e., the service along $\mathbb{P}_{f_4} (1)= (0, 4)$. However, this service depends on the burst of $f_3$ at the input of node 4, which depends in its turn on the minimum guaranteed service of $f_3$ until reaching the input of node 4, i.e., the service along $\mathbb{P}_{f_3} (1)= (0, 3)$. Detailing such dependencies for all the flows crossing the network reveals actually the need to quantify the service curve guaranteed to each flow $f$ along each of its subpaths, i.e., the service along $\mathbb{P}_{f} (n)$ for $\forall n \leq h$.

Expressing the service curves and the propagated bursts, for any flow along any of its subpaths, will define a system of linear equations. The latter can be solved using matrices, when a necessary and sufficient condition on the flow rates is verified. These different steps of our proposed PMOC approach, to compute the delay upper bounds, will be detailed in Sec. \ref{sec:service} and \ref{sec:computation}, and illustrated for a special case of ring networks in Sec. \ref{sec:SC}.

%###################################
\subsection{Service Curve for a Flow of Interest}
%###################################
\label{sec:service}
We focus herein on the first step of the PMOC approach, which consists in defining the guaranteed service curve for a \textit{f.o.i} along any of its subpaths in a ring network. We first present such a curve under arbitrary multiplexing within the crossed nodes in Th. \ref{Th:PMOO-Cycle}. Afterwards, we extend this result to FP multiplexing in Cor. \ref{Th:PMOO-Cycle-FP}. 
%%%%%%%%%%%%%%%%%Arbitrary Mux%%%%%%%%%%%%%%%%%%%%%%%%%%%%%%%%%%%
\begin{Theorem}
\label{Th:PMOO-Cycle}(Service Curve in Ring Networks under Arbitrary Multiplexing)
The service curve offered to a $f.o.i$ $f$ along its subpath, $\mathbb{P}_f (n)$, in a ring network under arbitrary multiplexing with strict service curve nodes of the rate-latency form $\beta_{R,T}$ and leaky bucket constrained arrival curves $\gamma_{\sigma, \rho}$, is a rate-latency curve, with a rate $R^{\mathbb{P}_f (n)}$ and a latency $T^{\mathbb{P}_f (n)}$,  defined as follows:
%========================================================================================================================================================
\begin{subequations}
\label{PMOO-service}
\begin{align}
&  R^{\mathbb{P}_f(n)} = \min \limits_{k \in \mathbb{P}_f(n)} [ R^{k}- \sum \limits_{i \ni k, i \neq f }{\rho_i} ] \label{PMOO-serviceR}\\
& T^{\mathbb{P}_f(n)} =  \sum\limits_{k \in \mathbb{P}_f(n)}T^{k}+  \sum\limits_{i \in \mathbb{K}_f(n)} \frac{\sigma_i^{0}.1_{\{f \ni i.ft\}} + \rho_i \cdot \sum\limits_{k \in\mathbb{P}_f(n)\cap \mathbb{P}_i} T^{k}} {R^{\mathbb{P}_f(n)}} \nonumber\\
& + \sum\limits_{i \in \mathbb{K}_f(n)} \frac{\sigma_i^{f.ft\ominus 1}.1_{\{i\ni f.ft  / i.ft \neq f.ft \}}} {R^{\mathbb{P}_f(n)}} \label{PMOO-serviceT} 
\end{align}
\end{subequations}
where $1_{\{cdt\}}$ is equal to $1$ if cdt is true and zero otherwise. 
\end{Theorem}

The proof of Th. \ref{Th:PMOO-Cycle} is provided in appendix \ref{proof}. As shown in Eq. (\ref{PMOO-serviceT}), some flow bursts are payed twice. These particular flows have actually two convergence points with the \textit{f.o.i}: their own source and the \textit{f.o.i} source; thus respecting the principle of the PMOC approach introduced in Sec. \ref{sec:PMOCbasics}.\\

\textit{We detail here the end-to-end service curve of the \textit{f.o.i} $f_1$ in the example of Fig. \ref{figure:cycle}, when the assumptions of the system model detailed in Sec. \ref{Model} are fulfilled, and all the crossed nodes offer the same service curve $\beta_{R,T}$. According to Th. \ref{Th:PMOO-Cycle}, the service curve of $f_1$ is a rate-latency curve, with a rate $R^{\mathbb{P}_{f_1}(3)}=\min[R-\rho_3-\rho_4,R-\rho_2, R- \rho_3]$ and a latency $T^{\mathbb{P}_{f_1}(3)}=3.T+\frac{1}{R^{\mathbb{P}_{f_1}(3)}}.( \sigma_2^0 +\rho_2.T + \sigma_3^0+ \rho_3.(2. T)+ \rho_4.T )+ \frac{1}{R^{\mathbb{P}_{f_1}(3)}}.( \sigma_3^4+\sigma_4^4)$.} \\

To extend such a result to the case of FP multiplexing, we need to introduce the following terms:
\begin{itemize}
\item $PL(i)$ for the priority level of flow $i$, where each crossed node has at maximum $NP$ priority levels and $0$ denotes the highest one;
\item $L_{max}(i)$ for the maximum packet length of flow $i$, integrating the communication protocol overhead;
\item $hp_f^k = \{ i \neq f/ i \ni k, PL(i) \leq PL(f) \}$ for the set of flows crossing the node $k$ excluding the \textit{f.o.i} $f$, with priority equal or higher than the $f$ one;
\item $lp_f^k = \{ i \ni k, PL(i) \geq PL(f) \}$ for the set of flows crossing the node $k$ with priority equal or lower than the $f$ one;
\item  $\mathbb{K}_{\leq f}(n) = \{i \neq f / \exists k \in \mathbb{P}_f(n) / i \ni k, PL(i) \leq PL(f) \}$ for the set of flows interfering with the \textit{f.o.i} $f$ along its subpath, $\mathbb{P}_f(n)$, with a priority equal or higher than $f$ one.
\end{itemize}

It is worth noting that the worst-case behavior under FP multiplexing is covered under Arbitrary multiplexing, but the latter may infer pessimistic bounds since it does not take into account the priority impact, i.e., any flow may be delayed by all the other flows independently from their priorities. Hence, to overcome such limitations, we define the guaranteed service curve for a \textit{f.o.i} in ring a network, under FP multiplexing, in Cor. \ref{Th:PMOO-Cycle-FP}.

\begin{Corollary}(Service Curve in Ring Networks under FP Multiplexing)
\label{Th:PMOO-Cycle-FP}
The service curve offered to a $f.o.i$ $f$ along its subpath, $\mathbb{P}_f(n)$, in a ring network under FP multiplexing with strict service curve nodes of the rate-latency type $\beta_{R,T}$ and leaky bucket constrained arrival curves $\gamma_{\sigma, \rho}$, is a rate-latency curve, with a rate $R^{\mathbb{P}_f (n)}$ and a latency $T^{\mathbb{P}_f (n)}$,  defined as follows:
%========================================================================================================================================================
\begin{equation}
\label{PMOO-service-FP}
\begin{array}{lll}
R^{\mathbb{P}_f(n)} &=& \min \limits_{k \in \mathbb{P}_f(n)} [ R^{k}- \sum \limits_{i \ni hp_f^k}{\rho_i} ] \\
T^{\mathbb{P}_f(n)} & =& \sum\limits_{k \in \mathbb{P}_f(n)}(T^{k} + \frac{\max_{i \in lp_f^k} L_{max}(i)}{R^k}) \\
&+ & \sum\limits_{i \in \mathbb{K}_{\leq f}(n)} \frac{\sigma_i^{0}.1_{\{f \ni i.ft\}}+ \rho_i \cdot \sum\limits_{k \in \mathbb{P}_f(n)\cap \mathbb{P}_i} (T^{k}+ \frac{\max_{j \in lp_f^k} L_{max}(j)}{R^k})} {R^{\mathbb{P}_f(n)}} \\
&+& \sum\limits_{i \in \mathbb{K}_{\leq f}(n)} \frac{\sigma_i^{f.ft\ominus 1}.1_{\{ i\ni f.ft / i.ft \neq f.ft\}}} {R^{\mathbb{P}_f(n)}} \\
\end{array} 
\end{equation}
\end{Corollary}

\proof
\label{proof:cor-FP}
The proof is straightforward following the Th. \ref{Th:PMOO-Cycle}. Under FP multiplexing, within each crossed node, a \textit{f.o.i} $f$ is selected for transmission only if all flows with equal or higher priorities are already transmitted. Furthermore, since the transmission is non-preemptive, $f$ may be blocked at the worst-case during the transmission time of one maximum packet length with a lower priority level.

Hence, we start by taking into account only the impact of lower priority flows on the \textit{f.o.i}, due to the non-preemptive transmission. The left-over service curve of each crossed node under FP is computed in this case through the application of Cor. \ref{cor:residual-sc}. The obtained service curve for each crossed node $k$ is a strict service curve and still has a rate-latency form, with a rate $R^k$ and a latency $\frac{\max_{j \in lp_f^k} L_{max}(j)}{R^k} + T^k$. Afterwards, we need to consider the impact of equal or higher priority flows in $\mathbb{K}_{\leq f}(n) $ when applying Th. \ref{Th:PMOO-Cycle}, to infer the guaranteed service curve of the \textit{f.o.i} $f$, which finishes the proof.
\endproof

\textit{We detail here the end-to-end service curve of the \textit{f.o.i} $f_1$ in the example of Fig. \ref{figure:cycle}. Consider that all the crossed nodes implement FP multiplexing with two priority levels and offer the same service curve $\beta_{R,T}$. Moreover, the flows $f_1$ and $f_3$ have the highest priority, whereas $f_2$ and $f_4$ have the lowest one. According to Cor. \ref{Th:PMOO-Cycle-FP}, the end-to-end service curve of the $f.o.i$ $f_1$ is a rate-latency curve, with a rate $R^{\mathbb{P}_{f_1}(3)}=\min[R-\rho_3,R, R- \rho_3]$ and a latency $T^{\mathbb{P}_{f_1}(3)}=3.T+ L_{max}(4)/R + L_{max}(2)/R +\frac{1}{R^{\mathbb{P}_{f_1}(3)}}.( \sigma_3^0+ \rho_3.(2.T+ L_{max}(4)/R) )+ \frac{1}{R^{\mathbb{P}_{f_1}(3)}}.\sigma_3^4$.}

%###################################
\subsection{Computation of the Delay Upper Bound}
%###################################
\label{sec:computation}
%definition of the system of linear equations for blind MUX
Now that we have expressed the service curve guarantees for each \textit{f.o.i} along any of its subpaths, we can move to the second step of the PMOC approach, which consists in computing the delay bounds. We put down all the system constraints in a ring network under arbitrary multiplexing, which depend on some variables, i.e., propagated bursts and the offered services:
\begin{itemize}
\item \textbf{Service Curve Constraint}\\
$\forall f \in I$, $\forall n \leq h$, for any  $]s,t]$, according to Th. \ref{Th:PMOO-Cycle},
\[ D_f^{f.ft\oplus(n-1)} (t) - A_f^{f.ft}(s) \leq \beta_{R^{\mathbb{P}_f (n)}, T^{\mathbb{P}_f (n)}} (t-s) \]

\item \textbf{Output Arrival Curve Constraint}\\
$\forall f \in I$, $\forall n \leq h$, according to Th. \ref{TH1},
\[\alpha_f^{f.ft\oplus(n-1)} (t) =\alpha^0 \oslash \beta_{R^{\mathbb{P}_f (n)}, T^{\mathbb{P}_f (n)}} (t)\]

\item \textbf{Delay bound}\\
$\forall f \in I$, $\forall n \leq h$, according to Th. \ref{TH1}, the delay bound of flow $f$ along its subpath $\mathbb{P}_f(n)$
\[  EED_{f}^{\mathbb{P}_f(n)} =h(\alpha^0, \beta_{R^{\mathbb{P}_f (n)}, T^{\mathbb{P}_f (n)}} )\]
\end{itemize}

In the case of rate-latency service curves and leaky-bucket arrival curves, these system constraints are linear and can be replaced with the following set (*):\\

\begin{itemize}
\item \textbf{Service Curve Constraint}\\
$\forall f \in I$, $\forall n \leq h$, for any  $]s,t]$,

\begin{subequations}
\begin{align}
&  R^{\mathbb{P}_f(n)} = \min \limits_{k \in \mathbb{P}_f(n)} [ R^{k}- \sum \limits_{i \ni k, i \neq f }{\rho_i} ] \nonumber\\
& T^{\mathbb{P}_f(n)} =  \sum\limits_{k \in \mathbb{P}_f(n)}T^{k}+  \sum\limits_{i \in \mathbb{K}_f(n)} \frac{\sigma_i^{0}.1_{\{f \ni i.ft\}} + \rho_i \cdot \sum\limits_{k \in\mathbb{P}_f(n)\cap \mathbb{P}_i} T^{k}} {R^{\mathbb{P}_f(n)}} \nonumber\\
& + \sum\limits_{i \in \mathbb{K}_f(n)} \frac{\sigma_i^{f.ft\ominus 1}.1_{\{i\ni f.ft  / i.ft \neq f.ft \}}} {R^{\mathbb{P}_f(n)}} \nonumber
\end{align}
\end{subequations}

\item \textbf{Output Arrival Curve Constraint}\\
$\forall f \in I$, $\forall n \leq h$, 
\[ \sigma_{f}^{f.ft \oplus (n-1)} =   \sigma_f^0 + \rho_f \times  T^{\mathbb{P}_f(n)} \]

\item \textbf{Delay bound}\\
$\forall f \in I$, $\forall n \leq h$, 
\[   EED_{f}^{\mathbb{P}_f(n)} = \frac{\sigma_{f}^{0} }{R^{\mathbb{P}_f(n)}} + T^{\mathbb{P}_f(n)}  \]
\end{itemize}

Hence, the set (*) can be written in a matrix form as follows:\\

\begin{itemize}
\item \textbf{Service Curve Constraint}
%=======================================
\begin{equation*}
\overbrace{\begin{bmatrix} 
 T^{\mathbb{P}_f(1)}\\ \vdots \\T^{\mathbb{P}_f(h_f)}\\ \vdots 
\end{bmatrix}}^T=  
\overbrace{\begin{bmatrix} 
c1_{f1}\\ \vdots \\ c1_{f h_f}\\ \vdots 
\end{bmatrix}}^{C1} +
\overbrace{\begin{bmatrix}
a1_{f,1}&\cdots&a1_{f, h_f} &\cdots\\
\vdots &\ddots&\ddots \\
a1_{f h_f ,1}&\cdots& \cdots & \cdots \\
\vdots&\vdots &\ddots&\ddots 
\end{bmatrix}}^{A1} \times 
\overbrace{\begin{bmatrix}
 \sigma_f^{f.ft \oplus 1}  \\ \vdots \\\sigma_f^{f.ft \oplus (h_f-1)}  \\ \vdots \\
\end{bmatrix}}^\sigma 
\label{formula:T}
\end{equation*}
%===============
where $T$ is the vector that holds the latencies of the offered service (Eq. (\ref{PMOO-serviceT})), $A1$ is the matrix of the coefficients of unknown propagated bursts and $C1$ is the vector of constants, i.e, the latencies $T^i$ and initial bursts transmission times, appearing in the service curve constraints of (*).\\
\item \textbf{Output Arrival Curve Constraint}
%=======================================
\begin{equation*}
\overbrace{\begin{bmatrix} 
 \sigma_f^{f.ft \oplus 1} \\ \vdots \\\sigma_f^{f.ft \oplus (h_f-1)}  \\ \vdots \\
\end{bmatrix}}^\sigma=  
\overbrace{\begin{bmatrix} 
c2_{f1}\\ \vdots \\ c2_{f h_f}\\ \vdots 
\end{bmatrix}}^{C2} +
\overbrace{\begin{bmatrix}
a2_{f,1} &\cdots&a2_{f, h_f} &\cdots\\
\vdots &\ddots&\ddots \\
a2_{f h_f ,1}&\cdots&\cdots &\cdots \\
\vdots&\vdots &\ddots&\ddots 
\end {bmatrix}}^{A2} \times 
\overbrace{\begin{bmatrix}
T^{\mathbb{P}_f(1)}\\ \vdots \\T^{\mathbb{P}_f(h_f)}\\ \vdots 
\end{bmatrix}}^T 
\label{formula:b}
\end{equation*}
%===============
where $\sigma$ is the vector of the unknown propagated bursts, $A2$ is the matrix of the coefficients of the corresponding unknown offered service latencies, i.e., the flow rate, and $C2$ is the vector of constants, i.e., the initial bursts $\sigma_f^0$, appearing in the output arrival curve constraints of (*). \\
\item \textbf{Delay bound}
%=======================================
\begin{equation*}
\overbrace{\begin{bmatrix} 
EED^{\mathbb{P}_f(1)} \\ \vdots \\EED^{\mathbb{P}_f(h_f)}\\ \vdots 
\end{bmatrix}}^{EED}=  
\overbrace{\begin{bmatrix} 
c3_{f1}\\ \vdots \\ c3_{f h_f}\\ \vdots 
\end{bmatrix}}^{C3} +
\overbrace{\begin{bmatrix}
T^{\mathbb{P}_f(1)}\\ \vdots \\T^{\mathbb{P}_f(h_f)}\\ \vdots 
\end{bmatrix}}^T 
\label{formula:EED}
\end{equation*}
%===============
where $C3$ is the vector of constants, i.e., the initial bursts transmission times, appearing in the delay bound constraints of (*). 
\end{itemize}

When propagating the different constraints, this matrix form is transformed to the following ($\mathbb{M}^*$):
%=======================================
\begin{equation}
\left\{ \begin{array}{c}
	  (Id - A1\times A2) \times T = C1 + A1 \times C2\\
	  EED = C3+ T
\end{array} 
\right.
\label{formula:PMOOmatrix}
\end{equation}
%==============================================

%property of delay bound existence if and only if the matrix XX is invertible for blind

Based on the matrix form $\mathbb{M}^*$, we deduce in the following corollary a necessary and sufficient condition on the existence of delay upper bounds for each \textit{f.o.i} along any of its subpaths, in ring networks under arbitrary multiplexing. This condition will be detailed in the next section for a special case of ring networks.\\

\begin{Corollary}(Delay Bound under Arbitrary Multiplexing)
\label{Th:EUD}
In a ring network under arbitrary multiplexing, the delay upper bound of each \textit{f.o.i} $f$ along its subpath $\mathbb{P}_f(n)$ exists and is at most equal to
\[ EED_{f}^{\mathbb{P}_f(n)} = \frac{\sigma_{f}^{0} }{R^{\mathbb{P}_f(n)}} + T^{\mathbb{P}_f(n)}  \]
if and only if the matrix  $(Id - A1\times A2)$ in $\mathbb{M}^*$ is invertible, i.e., its determinant is not zero.
\end{Corollary}

\begin{proof}
Based on known results in linear algebra, we can see from $\mathbb{M}^*$ that the vector of latencies $T$ exists and is unique, if and only if the square matrix $(Id - A1\times A2)$ is invertible. Under this necessary and sufficient condition, we have $T =(Id-A_1\times A_2)^{-1}\times (C1 + A1 \times A2)$. Consequently, $EED = C3+ (Id-A_1\times A_2)^{-1}\times (C1 + A1 \times A2)$ exists and is unique. This finishes the proof of Cor. \ref{Th:EUD}.
\end{proof}

%extension to FP policy with the iterative algorithm
Such a result is extended as follows under FP multiplexing. We need to order the delay bound computation according to the decreasing order of priority levels, i.e., computing the delay bounds of the highest priority first. We distinguish the following main steps:
\begin{enumerate}
\item For each priority level $p \in [0, NP-1]$, we define the corresponding matrix form $\mathbb{M}^*$, when including only the constraints related to the flows with equal or higher priority than $p$, i.e., $\forall f \in I$ with $PL(f) \leq p$. It is worth noting that the impact of lower priority flows is already integrated within the service curve formula, defined in Cor. \ref{Th:PMOO-Cycle-FP};
\item If the necessary and sufficient condition of Cor. \ref{Th:EUD} is satisfied, then we compute the delay bounds of all the flows of priority level $p$ along their subpaths;
\item The unknown parameters in $\mathbb{M}^*$ defined for the priority level $p$, i.e., propagated bursts and service latencies, are updated with the computed values in step 2;
\item If $p < NP-1$, then back to the step 1 when focusing on the priority level $p \leftarrow p+1$.
\end{enumerate}

Hence, we have the following corollary concerning the computed delay bounds for each \textit{f.o.i} of priority level $p$ along any of its subpaths, in ring networks under FP multiplexing:
\begin{Corollary}(Delay Bound under FP Multiplexing)
\label{Th:EUD-FP}
In a ring network under FP multiplexing, the delay upper bound of each \textit{f.o.i} $f$ of priority level $p$ along its subpath $\mathbb{P}_f(n)$ exists and is at most equal to
\[ EED_{f}^{\mathbb{P}_f(n)} = \frac{\sigma_{f}^{0} }{R^{\mathbb{P}_f(n)}} + T^{\mathbb{P}_f(n)}  \]
if and only if for each priority level $pp$ higher than $p$, the matrix $(Id - A1\times A2)$ in $\mathbb{M}^*$ associated to the priority level $pp$ is invertible, i.e., its determinant is not zero.
\end{Corollary}

\begin{proof}
The proof is straightforward following the Cor. \ref{Th:EUD}. Following the main steps of the delay bound computation under FP multiplexing, detailed above, we have to verify in step 2 the necessary and sufficient condition of Cor. \ref{Th:EUD} for each priority level $pp$ higher than the \textit{f.o.i} priority level $p$, which finishes the proof. 
\end{proof}

%###################################
\subsection{Special Case: Regular Ring Networks}
%###################################
\label{sec:SC}
We introduce herein a particular case of ring networks, called regular ring networks, for which we deduce a specific necessary and sufficient condition for the existence of delay upper bounds, in comparison to the general one in Cor. \ref{Th:EUD}.

 %========================================================================================================================
\begin{Definition}(Regular Ring Network)
\label{RRN-def}
A ring network connecting $M$ nodes is a regular ring network with a degree $h$, where $ 2 \leq h \leq M $, when it satisfies the following assumptions: (i) all the nodes guarantee the same rate-latency service curve, $\beta_{R,T}$ and implement arbitrary multiplexing; (ii) each node $l \in [1, M]$ is generating a $(\sigma, \rho)$-constrained flow, destined to all its $k$-th downstream nodes from $l$, $\forall k \leq h$.
\end{Definition}
%===============================

It is worth noting that a ring network with a broadcast communication pattern is a regular ring network with a degree $h=M$. 

We have the following conjecture on the delay bounds in regular ring networks, based on a more specific necessary and sufficient condition than the one in Cor. \ref{Th:EUD}:
%========================================================================================================================
\begin{Conjecture}(Delay Bound in Regular Ring Networks)
\label{NSC-regular-def}
In a regular ring network under arbitrary multiplexing and with a degree $h$, the delay upper bound of each \textit{f.o.i} $f$ along its subpath $\mathbb{P}_f(n)$ exists and is at most equal to
\[ EED_{f}^{\mathbb{P}_f(n)} = \frac{\sigma_{f}^{0} }{R^{\mathbb{P}_f(n)}} + T^{\mathbb{P}_f(n)}  \]
if and only if the following equivalent conditions are verified:\\
%\begin{itemize}
(i) (Flow rate Cdt.) The maximum rate of each generated $(\sigma, \rho)$-constrained flow is as follows: $\rho < \frac{R}{2\cdot(h-1)}$;\\
(ii) (Utilization rate Cdt.) The maximum utilization rate of the network, $U_{max} = h\cdot\rho/R$, is as follows: $U_{max} < \frac{h}{2\cdot(h-1)}$. Thus, as $h \to \infty$, the maximum utilization rate tends to $50$\%.
%\end{itemize}
\end{Conjecture}
%===============================

\begin{figure}[htbp]
\centering
\includegraphics[scale=0.6]{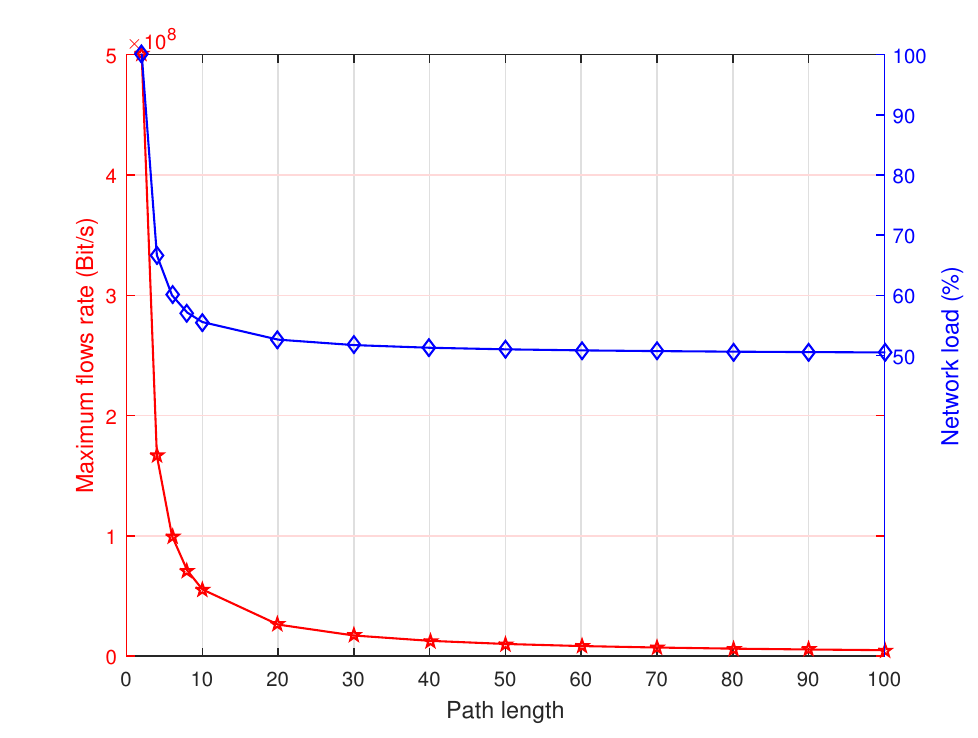}
\caption{Maximum network utilization and flow rate vs network degree, i.e., flow path length, for which the determinant of the matrix $(Id - A1\times A2)$ in $\mathbb{M}^*$ vanishes}
\label{fig:load}
\end{figure}

This conjecture is based on the observation of the behavior of the maximum utilization rate (resp. maximum flow rate), satisfying the necessary and sufficient condition of Cor. \ref{Th:EUD}, for regular ring networks when varying the degree $h \in [2, M]$, as illustrated in Fig. \ref{fig:load}. We actually have built the associated matrix form $\mathbb{M}^*$ for $h \in [2, 100]$ and $R= 1Gb/s$. Then, based on a symbolic computation tool, we have computed the maximum utilization rate of the network (resp. maximum flow rate), for which the determinant of the matrix $(Id - A1\times A2)$ in $\mathbb{M}^*$ vanishes. As we can see, The maximum network utilization rate decreases from $100\%$ for $h=2$ to $50.5\%$ for $h=100$, while the maximum flow rate decreases from $\frac{R}{2}$ for $h=2$ to $\frac{R}{198}$ for $h=100$. These values are coherent with the flow rate and network rate condition defined in the Conjecture \ref{NSC-regular-def}, which specify a maximum network utilization rate of $\frac{h}{2\cdot(h-1)}$ and a the maximum flow rate of $\frac{R}{2\cdot(h-1)}$. It is worth noting that the maximum utilization rate in Conjecture \ref{NSC-regular-def} is more restrictive than the one in Sec. \ref{Model}, i.e., $h\cdot\rho/ R \leq 1$.
%================================================================================
%================================================================================

\begin{figure}[htbp]
\centering
\includegraphics[scale=1.1]{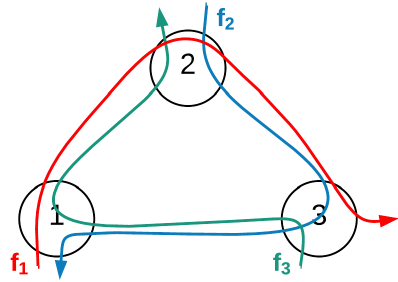}
\caption{Example of a regular ring network with $M=3$ and $h=2$}
\label{fig:example33Bis}
\end{figure}

%example
\textbf{Example}\\
We now explicit the matrix form $\mathbb{M}^*$ and the necessary and sufficient condition on the existence of delay bounds for the example illustrated in Fig. \ref{fig:example33Bis}. The latter is a regular ring network with 3 nodes, labeled from 1 to 3, and a degree $h=2$. Each node $i$ sends a $(\sigma^0,\rho)$-constrained flow $f_i$ and guarantees a service curve $\beta_{R,0}$. The aim is to compute the end-to-end delay bound of the \textit{f.o.i} $f_1$.
%================================================================================
%================================================================================

First, we explicit the different parameters of the matrix form $(*)$ in Sec. \ref{sec:computation} of such a network as follows:\\

\[T^T = (T^{\mathbb{P}_{f_1(1)}}, T^{\mathbb{P}_{f_1(2)}}, T^{\mathbb{P}_{f_2(1)}},T^{\mathbb{P}_{f_2(2)}},T^{\mathbb{P}_{f_3(1)}}, T^{\mathbb{P}_{f_3(2)}})\]

%%%%%%%%%%%%%%%%%%%%%%%%%%%%%
\[C1^T=\frac{\sigma^0}{R-\rho} \cdot (0,1, 0, 1, 0, 1)\]

%%%%%%%%%%%%%%%%%%%%%%%%%%%%%
\[
A_1=\frac{1}{R-\rho}\cdot\left(
\begin{array}{cccccc}
0&0&0&0&1&0 \\
0&0&0&0&1&0 \\
1&0&0&0&0&0 \\
1&0&0&0&0&0 \\
0&0&1&0&0&0 \\
0&0&1&0&0&0  
\end{array}\right)
\]

%%%%%%%%%%%%%%%%%%%%%%%%%%%%%
\[\sigma^T=(\sigma_{f_1}^1, \sigma_{f_1}^2, \sigma_{f_2}^2, \sigma_{f_2}^3, \sigma_{f_3}^3, \sigma_{f_3}^1)\] 
\[ C2^T=\sigma^0\cdot (1, 1, 1, 1, 1, 1) \] 
\[ C3^T=\frac{\sigma^0}{R-\rho}\cdot (1, 1, 1, 1, 1, 1)\]
\[ A2=\rho\cdot I_{(h\times M)} \]

Then, to verify the necessary and sufficient condition defined in Cor. \ref{Th:EUD}, we express the determinant of the matrix $(Id-A_1\times A_2)$, which is as follows:
\[(\rho-\frac{R}{2})\cdot(-2\rho^2+2R\rho -2R^2)/(R-\rho)^{3}\]

This function vanishes for the maximum flow rate $\rho=\frac{R}{2}$. This value is coherent with the Conjecture \ref{NSC-regular-def}, where the upper bound of the maximum flow rate is $ < R/2\cdot(h-1)$, i.e., $R/2$ for $h=2$. Hence, if the flow rate condition is verified, i.e., $\rho < R/2$, then the end-to-end delay upper bound of the \textit{f.o.i} $f_1$, $EED_{f_1}^{\mathbb{P}_{f_1}(2)}$, exists and is at most equal to $\frac{\sigma^0 }{R^{\mathbb{P}_{f_1}(2)}} + T^{\mathbb{P}_{f_1}(2)}$, where $R^{\mathbb{P}_{f_1}(2)}=R-\rho$ and $T^{\mathbb{P}_{f_1}(2)} = \frac{2\sigma^0}{R-\rho} +\frac{\sigma^0\rho(\rho^2-R\rho+R^2)}{(R- \rho)(R^3 -3R^2\rho -2\rho^3)}$

%=====================================================================================================================================
\subsection{Performance Evaluation}
\label{Evaluation}
In this section, we detail some numerical results of the delay upper bounds of a \textit{f.o.i}. in a ring network with cyclic dependencies, under different scenarios, when applying our approach PMOC. First, we describe the considered case study and scenarios. Then, we report the sensitivity analysis of such computed upper bounds with respect to flows burst, rate and path length, for various values of network size $M$. Finally, we assess their tightness in several scenarios, in reference to a lower bound on WCD (Worst-Case Delay). 

\subsubsection{Case study and scenarios}
We consider the case study with the following assumptions:
\begin{itemize}
	\item The topology is a unidirectional ring topology, connecting $M$ nodes;
	\item All nodes guarantee a rate-latency service curve $\beta_{R,T}$ with $ R= 1Gbps$ and $T= 600 ns$;
	\item Each node generates one leaky-bucket constrained flow with a burst $\sigma$ and a rate $\rho$;
	\item The considered network is a regular ring network with a degree $h$, according to Def. \ref{RRN-def}.
	\end{itemize}

To analyse the sensitivity of the computed delay bounds and to assess their tightness, we consider various network configurations, where each network configuration is defined with the tuple $(\sigma, \rho, h, M)$. The main idea is to vary only one parameter of this tuple at a time, to highlight its impact on the computed delay bounds.

\subsubsection{Sensitivity analysis}
\begin{figure*}[htbp]
\centering
\includegraphics[scale=0.5]{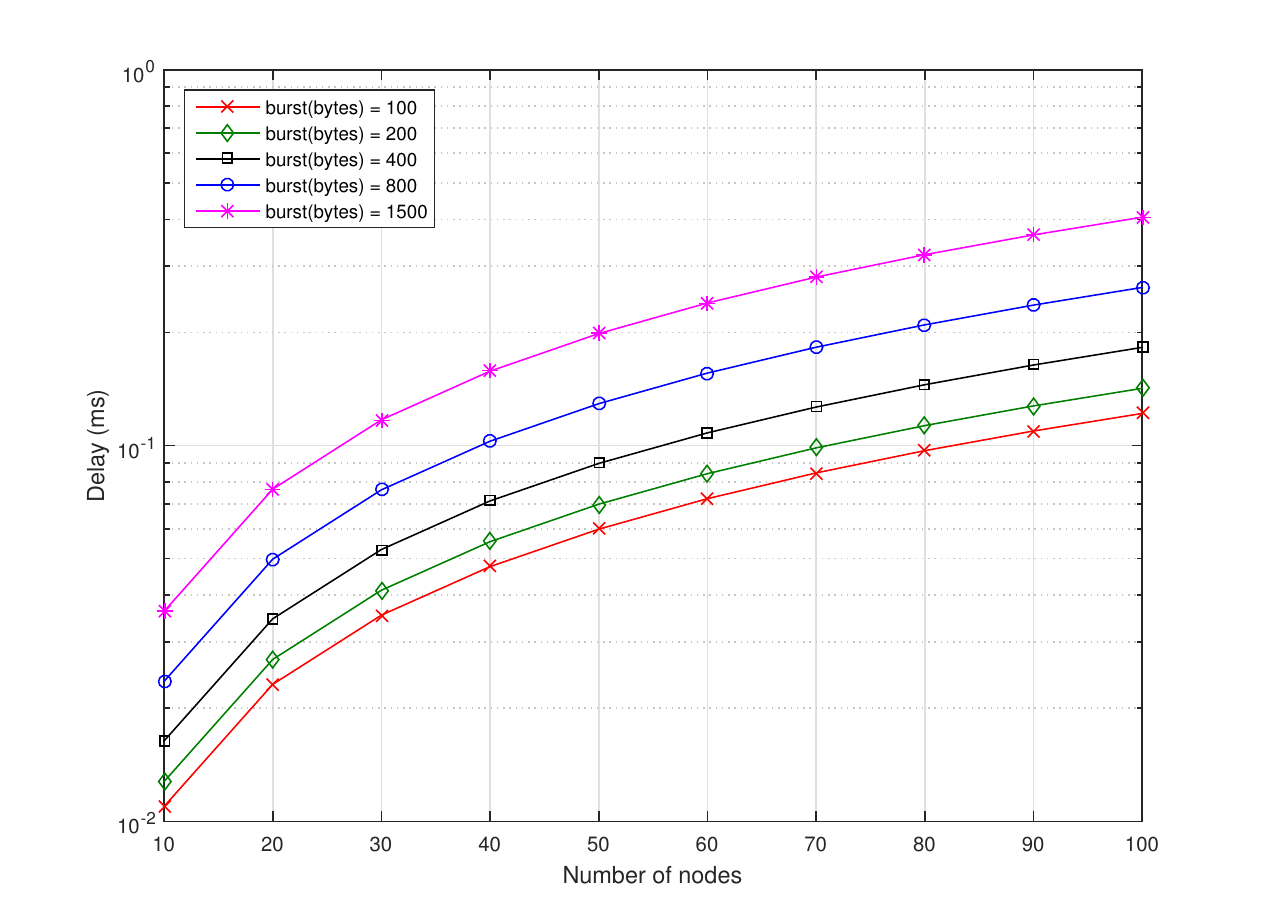}
\caption{The impact of the flow burst on the delay bounds vs network size for $(\sigma \in [100-1500]bytes, \rho=128Kbps, h=M, M \in [10-100])$.}
\label{fig:sensitivity1a}
\end{figure*}

\begin{figure*}[htbp]
\centering
\includegraphics[scale=0.5]{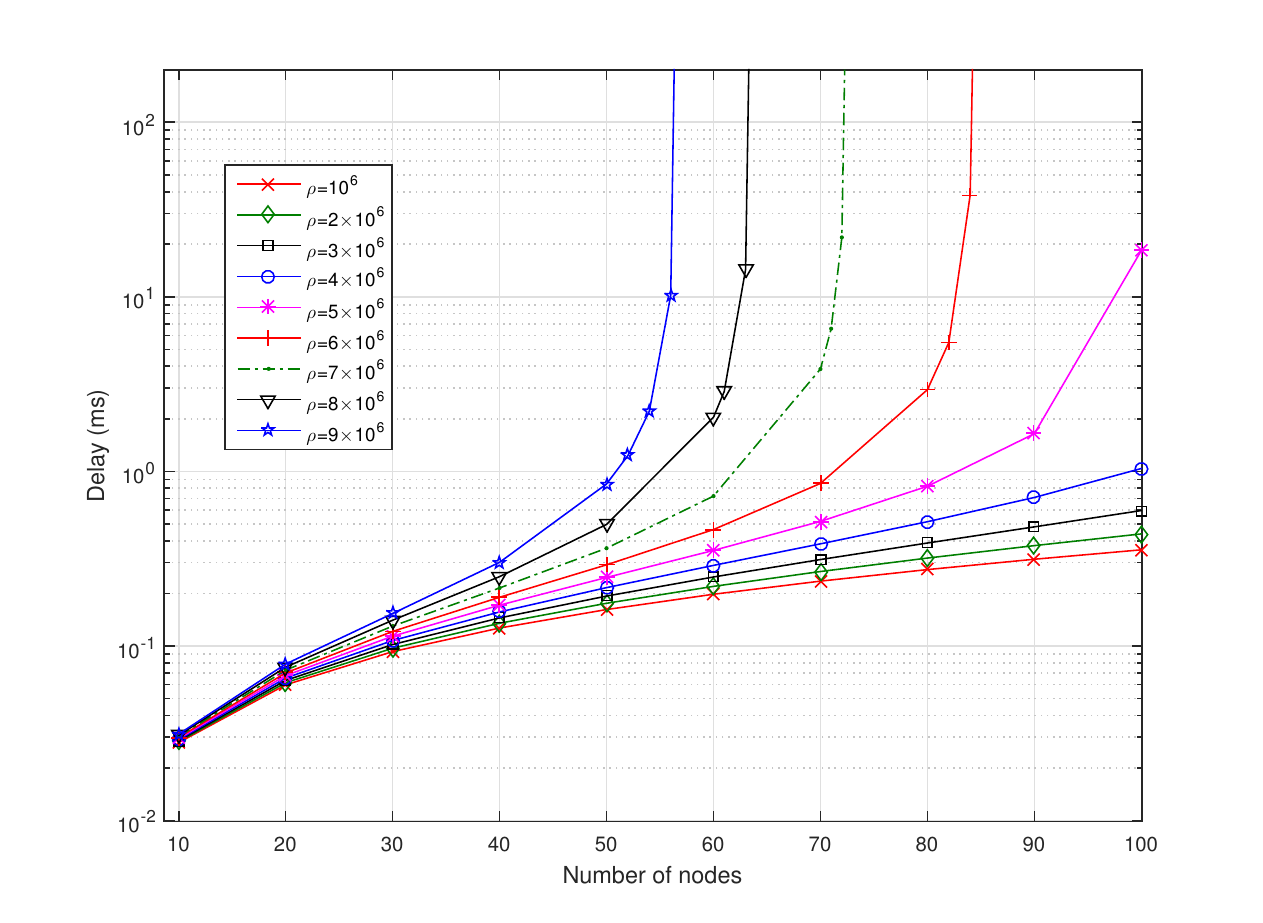}
\caption{The impact of flow rate on the delay bound vs network size for $(\sigma=128bytes, \rho=[1-9]Mbps, h=M, M \in [10-100])$.}
\label{fig:sensitivity1b}
\end{figure*}

\begin{figure*}[htbp]
\centering
\includegraphics[scale=0.75]{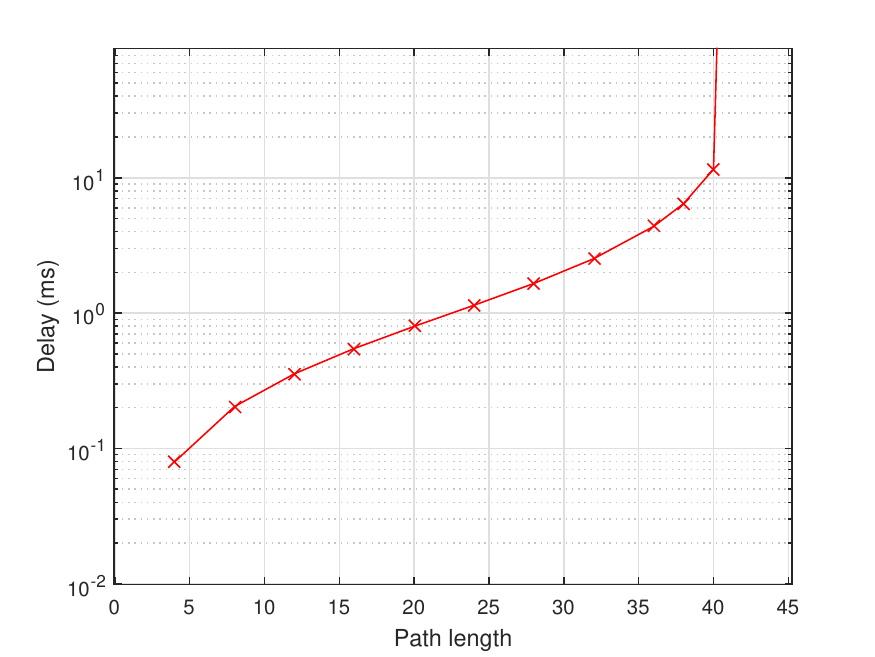}
\caption{The impact of the flow path on delay bound for $(\sigma=1500bytes, \rho=12Mbps, h\in[4-45], \forall M > h)$.}
\label{fig:sensitivity1c}
\end{figure*}

We discuss herein the impact of each network configuration parameter on the delay bounds, computed with the PMOC approach. The numerical results for different scenarios are reported in Figs. \ref{fig:sensitivity1a}, \ref{fig:sensitivity1b} and \ref{fig:sensitivity1c}.

Fig. \ref{fig:sensitivity1a} shows the impact of the burst size on the delay bounds. Obviously, for a fixed network size, the delay increases when increasing the flow burst, since the multiplexing time increases within each crossed node. Moreover, for a fixed flow burst, the delay increases with the network size. There are two main observations to note from this analysis scenario: 
\begin{itemize}
	\item the delay bound grows logarithmically in terms of flow burst, e.g., for $M=100$, when the flow burst increases from $100bytes$ to $1500bytes$, i.e., $\times15$, the delay goes only from $0.12ms$ to $0.4ms$, i.e., $\times 3.3$; 
	\item the delay bound for a fixed flow burst increases in a more noticeable way with the network size but still grows linearly, e.g., for $\sigma=100$bytes, the delay goes from $10^{-2}$ms for $M=10$ nodes to almost $10^{-1}$ms for $M=100$ nodes, i.e., $\times10$, which is equivalent to the scaling factor of the network.
\end{itemize}
These results infer that the interfering flow bursts have higher impact on the delay bound of a \textit{f.o.i} than its own burst. This fact is very coherent with the delay bound expression, defined in Sec. \ref{sec:computation}.

%======================================================================================
Fig. \ref{fig:sensitivity1b} shows the impact of the flow rate on the delay bounds. As we can notice, there are two distinguishable behaviors of the delay bounds: 
\begin{itemize}
	\item when the flow rate condition in Conjecture \ref{NSC-regular-def} is verified, the delay bounds grow logarithmically in terms of the flow rate, e.g., for $M=40$, when the rate increases from $1$ Mb/s to $9$Mb/s, i.e., $\times9$, the delay bound grows from almost $10^{-2}$ms to $3.10^{-2}$ms, i.e., $\times3$; 
	\item when this condition is violated, the delay bound tends to infinity, e.g., for $\rho=8Mb/s$, the delay bound diverges for a network size higher than $M=63$, which corresponds to the condition $\rho<\frac{R}{2(M-1)}\Leftrightarrow  M<\frac{R}{2\rho}+1=63.5$. This fact infers an exponential growth of the delay bounds with the network size, when the flow rate condition achieves its limit.
\end{itemize}
These results show the inherent impact of the flow rate on the delay bounds with the PMOC approach, which is relevant with our conjecture on the network stability condition of regular ring networks in Sec.\ref{sec:SC}.

%======================================================================================
Fig. \ref{fig:sensitivity1c}, shows the impact of the flow path length on the delay bounds. As it is shown, the delay bound has similar behavior in terms of flow path length than its rate, i.e., grows logarithmically when the flow rate condition is verified. Increasing the flow path length induces a higher number of interfering flows along the path; thus a higher service latency and lower service rate according to the PMOC approach. Moreover, it is worth noting that the delay bounds for regular ring networks depend only on the network degree $h$, i.e., flow path length. For instance, the delay bound is $0.79ms$ for $h=20$ independently from the network size. This result is coherent with Conjecture\ref{NSC-regular-def}.

\textbf{\textit{These results show that the delay bounds computed with the PMOC approach are particularly sensitive to the flow rate and path length. This fact is mainly due to the conditions in Conjecture \ref{NSC-regular-def}, which depend on both parameters and infer an exponential behavior of the delay bounds when they achieve their limit.}}

\subsubsection{Tightness analysis}
To investigate the tightness of our approach, we compare the delay bounds obtained with our proposed method to an achievable worst-case delay, denoted as WCD lower bound. The latter is computed when considering an intuitive worst-case scenario, which consists in integrating for each $f.o.i$ only the impact of downstream flows interferences within each crossed node, and ignoring the impact of the upstream flows at its source node, i.e., this is the unknown variable due to cyclic dependency and it is considered as null for this intuitive WCD. The size of the interval between the computed upper delay bounds and WCD lower bounds will give us an idea about the delay bound tightness, i.e., this interval includes the exact worst-case delay; thus if this interval duration is small, then the upper bound delay is tight.

\begin{figure*}[htbp]
\centering
\includegraphics[scale=0.5]{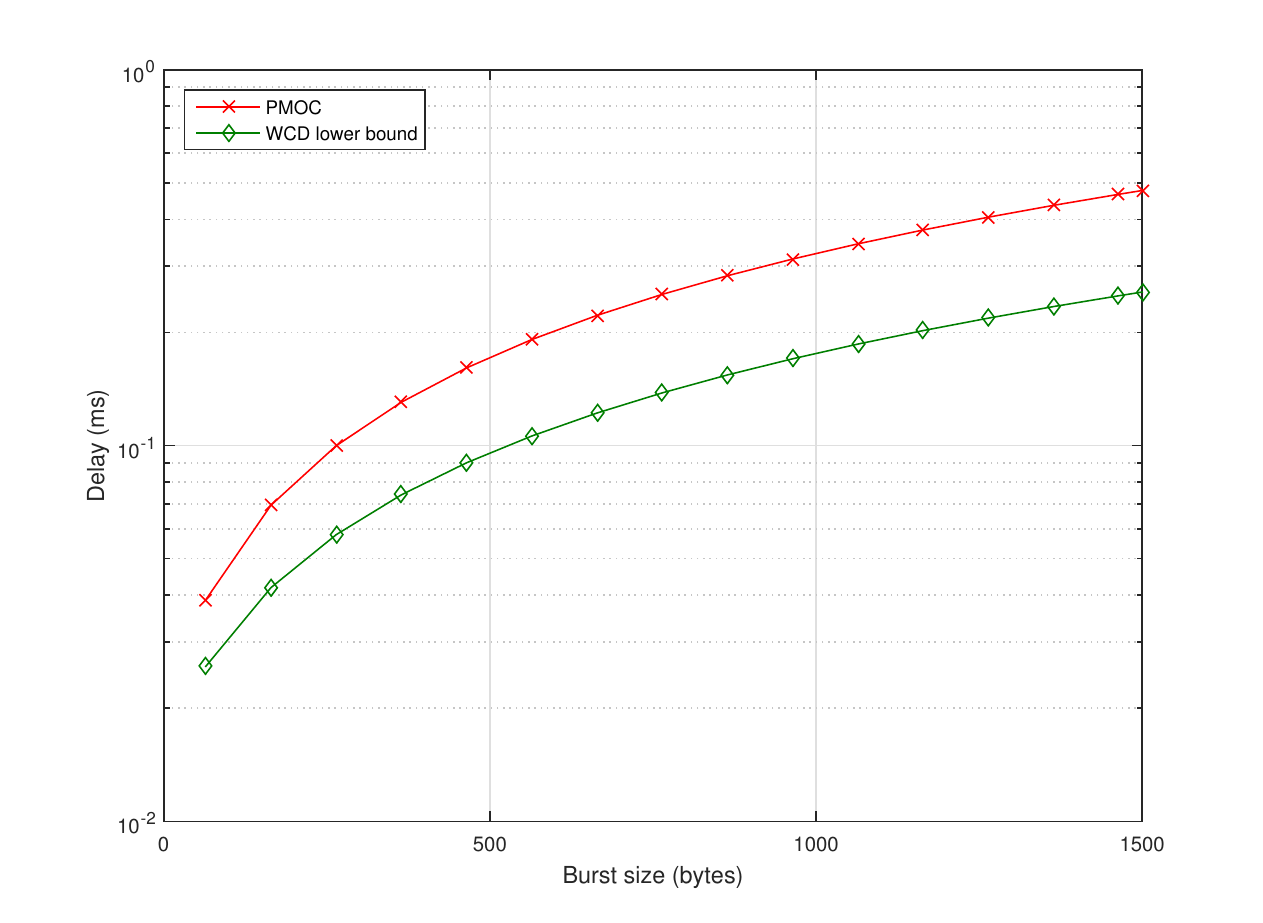}
\caption{Impact of the burst on delay bound tightness for $(\sigma=[64-1500]bytes, \rho=128Kbps, h=M, M=20)$.}
\label{fig:tightness1a}
\end{figure*}

\begin{figure*}[htbp]
\centering
\includegraphics[scale=0.5]{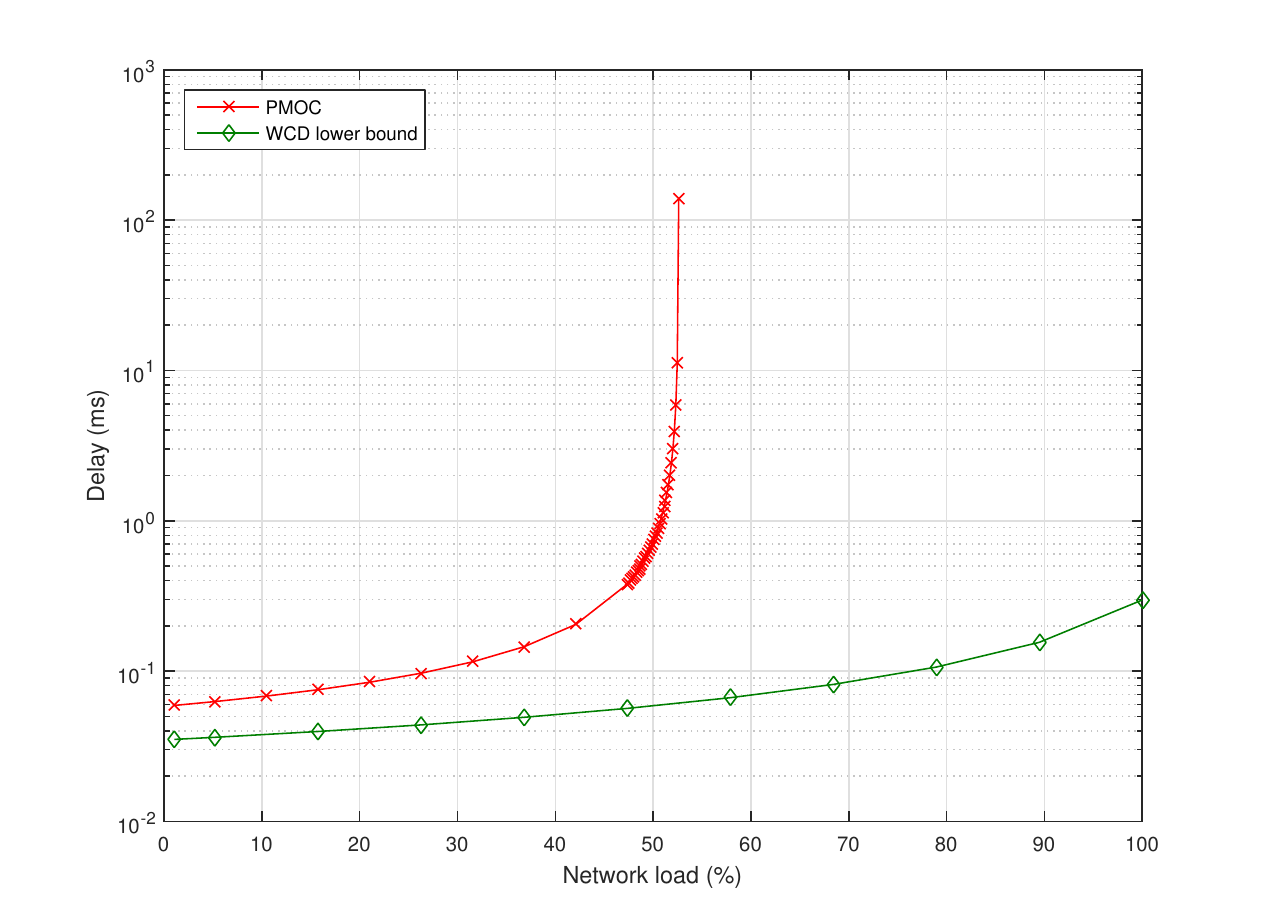}
\caption{Impact of the maximum network utilization rate on delay bound tightness for $(\sigma=128bytes, \rho=[0.5-50]Mbps, h=M, M=20)$.}
\label{fig:tightness1b}
\end{figure*}

\begin{figure*}[htbp]
\centering
\includegraphics[scale=0.5]{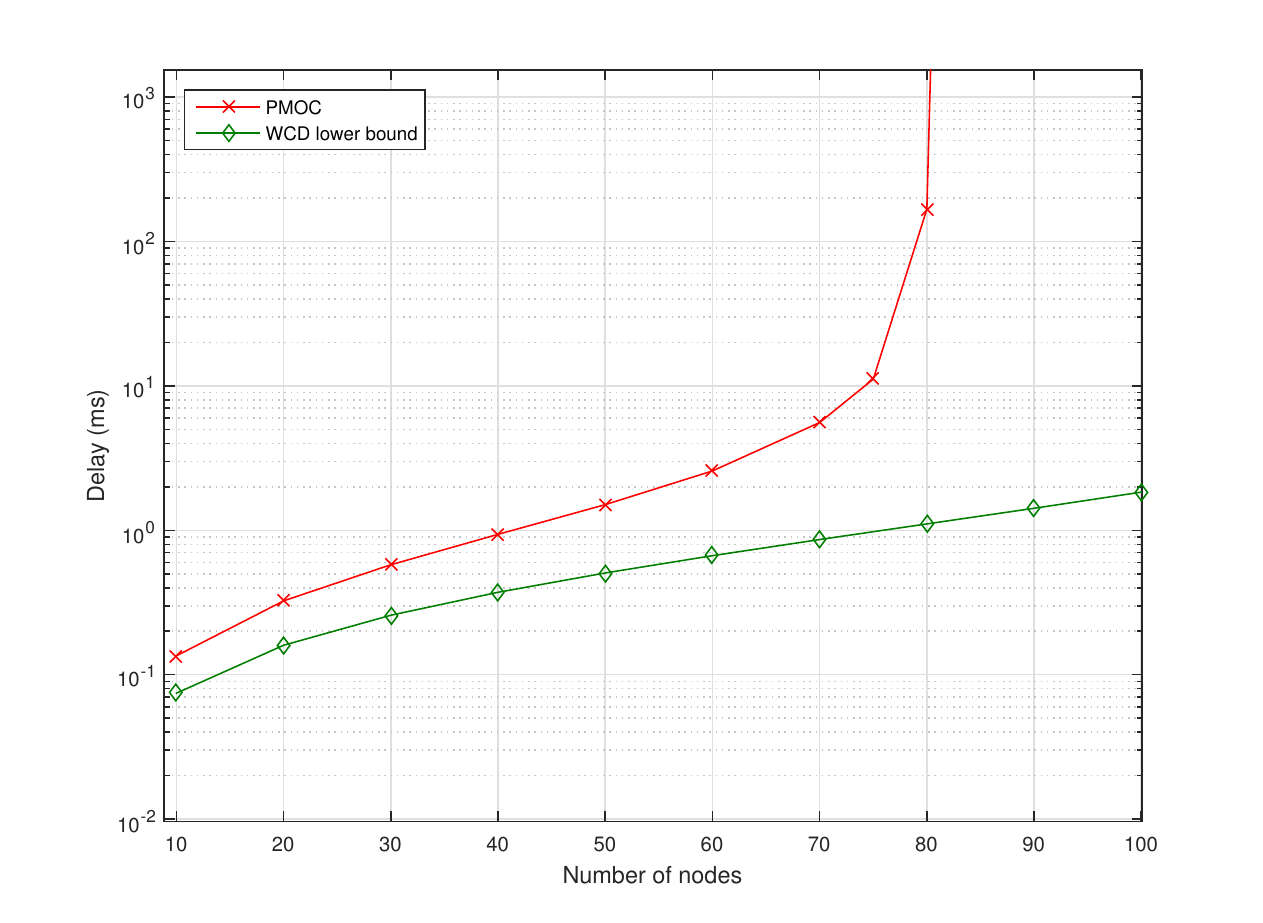}
\caption{Impact of network size on delay bound tightness for $(\sigma=787bytes, \rho=6.3Mbps, h=M, M \in [10-100])$.}
\label{fig:tightness1c}
\end{figure*}

Figs. \ref{fig:tightness1a}, \ref{fig:tightness1b} and \ref{fig:tightness1c} report the numerical results of different analysis scenarios, conducted to assess the delay bounds tightness. As we can notice, the gap between the delay bound computed with the PMOC approach and the WCD lower bound still is bounded and both curves have the same shape, when varying the flow burst (Fig. \ref{fig:tightness1a}), the network utilization rate (Fig. \ref{fig:tightness1b}) and flow path length, i.e., the network size for broadcast pattern, (Fig. \ref{fig:tightness1c}), if the network stability condition is verified, i.e., $U_{max}<\frac{h}{2(h-1)}=52.6\%$ for $M=100$.

However, when the network utilization rate condition is violated, we can not conclude on the delay bound tightness since it tends to infinity.

\textbf{\textit{These results show that: if the network utilization rate condition is verified, then the delay bounds computed with the PMOC approach have an acceptable tightness, when varying different network and flow parameters.}}

\subsubsection{Comparison with the Related Work}
In order to benchmark the delay bounds obtained with the PMOC approach against the existing ones, i.e., Time Stopping and Backlog-based, we consider the same case of study and scenario detailed in Sec. \ref{RW-results}.
\begin{figure*}[htbp]
		\centering
		\includegraphics[scale=0.45]{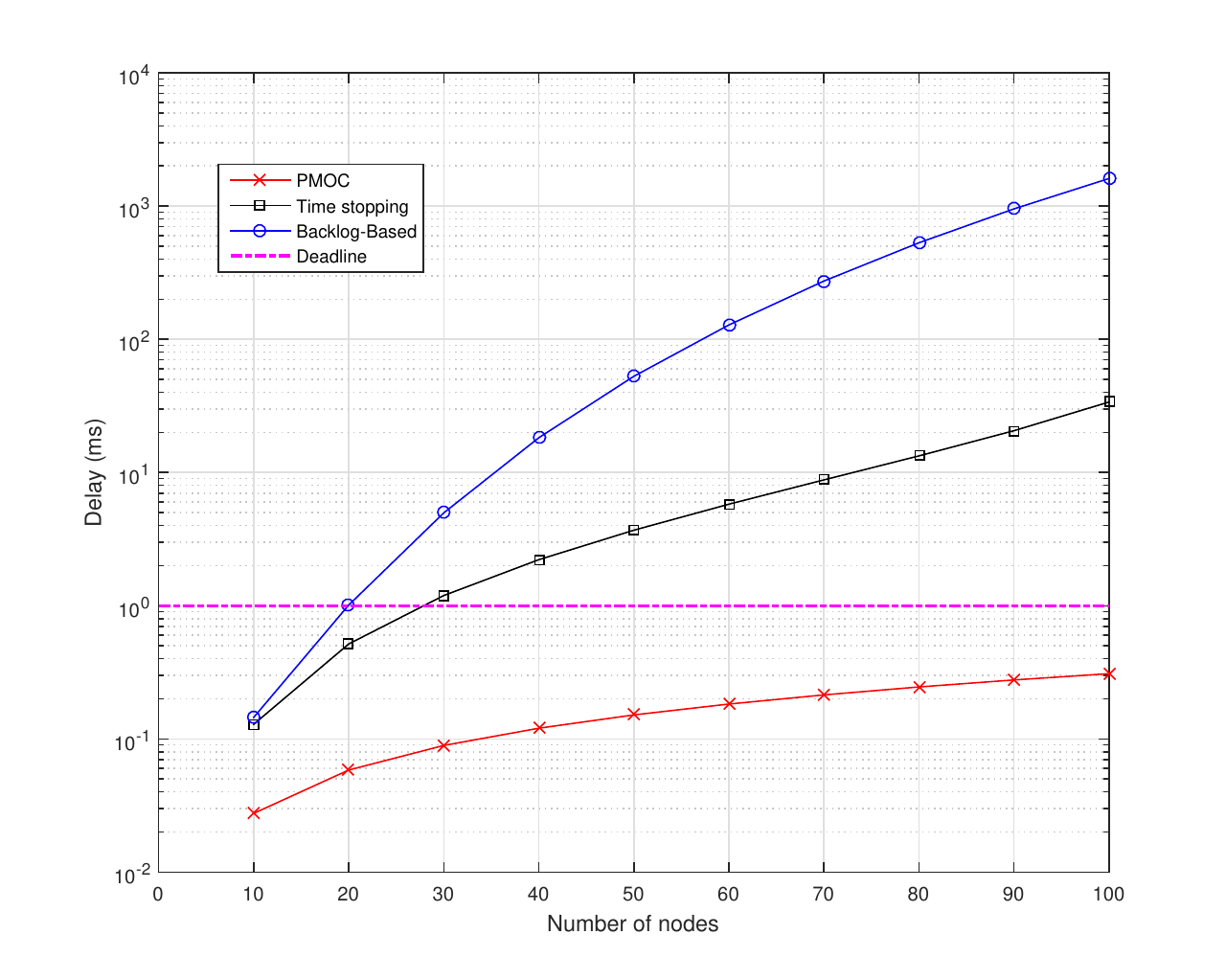}
	\caption{End-to-end delay bounds vs number of nodes for $(\sigma=128bytes, \rho=128Kbps,h=M, M \in [10-100])$.}
	\label{fig:RW2-comparisonb}
\end{figure*}

Fig. \ref{fig:RW2-comparisonb} shows a comparison of the different approaches when enlarging the network size. As we can notice, the PMOC approach offers tighter delay bounds for large-scale networks while guaranteeing the flows deadline, in comparison with the conventional methods, e.g., for a network of $100$ nodes, the PMOC delay is $0.3$ms compared to $33.8$ms and $1.6$s for Time-Stopping and Backlog-based methods, respectively. Hence, the maximum network size respecting the flow deadline is about 20 and 27 nodes with the Backlog-based and Time Stopping methods, respectively; whereas it achieves 100 nodes with PMOC approach. This represents an enhancement of network scalability up to $400$\% under PMOC, with reference to conventional timing analyses.
\begin{figure*}[htbp]
		\centering
		\includegraphics[scale=0.45]{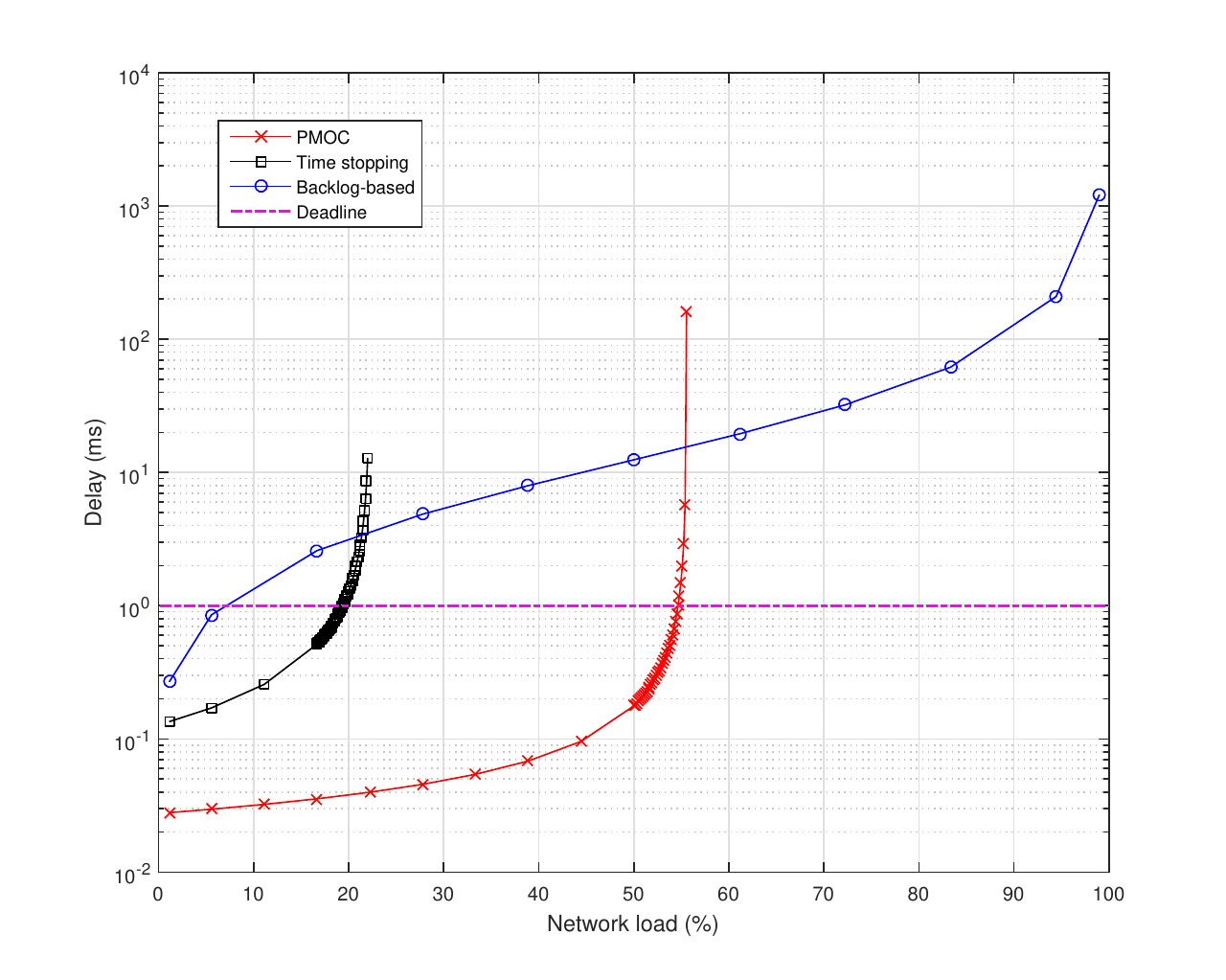}
	\caption{End-to-end delay bounds vs network utilization rate for $(\sigma=128bytes, \rho \in [1-100]Mbps, h=M, M =10)$.}
	\label{fig:RW2-comparisonc}
\end{figure*}

Fig. \ref{fig:RW2-comparisonc} illustrates the impact of increasing the congestion on the different methods. As we can see, the Time Stopping method diverges for a global utilization rate around 22.22\%, which corresponds to $\frac{2}{M-1}$ as explained in Sec. \ref{RW-results}; whereas it achieves 55.55\% with our proposed approach, which corresponds to the upper bound on the network utilization rate in Conjecture 1 when $h=M$: $\frac{M}{2(M-1)}$. However, a full utilization rate is still achievable under the Backlog-based method, even if the delay bounds are overly pessimistic, e.g., $1,22$s for $U_{max}=99\%$. Furthermore, the maximum network utilization rate respecting the flows deadline is only about $7.1\%$ and $19.36\%$ with the Backlog-based and Time Stopping methods, respectively, compared to $54.6\%$ with PMOC. This represents an enhancement of resource efficiency up to $670$\% under PMOC, with reference to conventional timing analyses.\\

\textbf{\textit{This comparative analysis shows that using PMOC approach yields enhanced network performance, in terms of resource efficiency and network scalability, in comparison with the conventional timing analyses.}}
%########################################################
%########################################################

%########################################################
%########################################################

\section{Generalization of PMOC for Multiple-Ring Networks}
\label{general-analysis}
We detail in this section the generalization of the PMOC approach to be applicable for the multiple-ring networks. First, we adapt the system model defined for mono-ring networks in Sec. \ref{Model} to fit the multiple-ring networks. Then, we define the guaranteed service curves for a \textit{f.o.i} along any of its subpaths for such a topology under Arbitrary multiplexing in Cor. \ref{th:GPMOC} and Fixed Priority multiplexing in Cor. \ref{cor:PMOC-FP}. Afterwards, we explicit the end-to-end delay bound computation for an illustrative example of a multiple-ring network, through detailing the corresponding matrix form $\mathbb{M}^*$ and the necessary and sufficient condition. Finally, we analyse the sensitivity of the derived delay bounds with respect to several network and flows parameters with reference to the mono-ring network.

\subsection{Extended System Model}
\label{Model-ext}
The system model of mono-ring networks in Sec. \ref{Model} still is applicable for multiple-ring networks, when considering the following adaptations:
%================================================================================
\begin{figure}[htbp]
\centering
\includegraphics[scale=0.8]{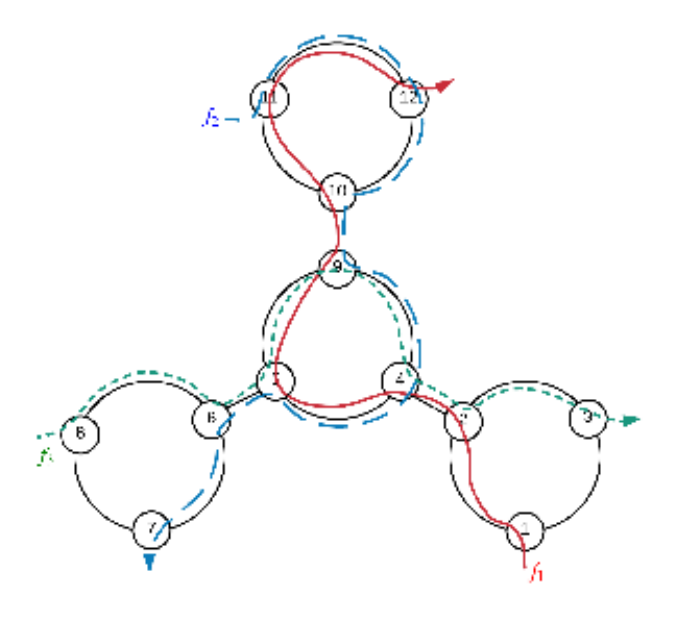}
\caption{Example of a multiple-ring with three flows}
\label{fig:exampleM}
\end{figure}
%================================================================================
%%%====================
\begin{itemize}
\item We consider a multiple-ring topology, as shown in Fig. \ref{fig:exampleM}, consisting of $nbR$ unidirectional rings, connecting $M$ nodes, and serving a fixed set of flows $I$. The key idea is to gather nodes in peripheral rings according to their exchanged data. The peripheral rings are connected to the backbone ring via specific nodes, called gateways, which manage the inter-ring communications and do not generate any flows, i.e., just forward. This fact may improve the utilization rate and the end-to-end delay within each peripheral ring, since it isolates the intra-ring traffic from the inter-ring one;

\item Each flow $i \in I$ follows a fixed path from its initial source until the final sink, defined as $\mathbb{P}_i= (0, i.ft, i.ft\oplus 1, ..., i.ft\oplus(h_i-1))$ similarly to the mono-ring case. However, the only differences are the definition of the notations $l \oplus k$ and $l \ominus k$, which designate the $k-th$ node downstream and upstream from node $l$ with reference to the path of flow $i$, respectively, e.g., the first upstream node for node $9$ along the path of flow $f_1$ is node $5$, as shown in Fig. \ref{fig:exampleM};

\item In multiple-ring network, a $f.o.i$ $f$ can have more than two convergence points with an interfering flow $i$ unlike the mono-ring case, e.g., at the $f.o.i$ source, at the interfering flow source and at the backbone. We denote $conv(i,f,n)$ the convergence points of the $f.o.i$ $f$ with the interfering flow $i$ along its subpath of length $n$ in a multiple-ring network. In the example of Fig. \ref{fig:exampleM}, the flows $f_1$, $f_2$ and $f_3$ have the following paths $\mathbb{P}_{f_1}$=\{0,1,2,4,5,9,10,11\}, $\mathbb{P}_{f_2}=\{0,11,12,10,9,4,5,6\}$ and $\mathbb{P}_{f_3}=\{0,8,6,5,9,4,2\}$, respectively. The convergence points between the $f.o.i$ $f_1$ and $f_2$ (resp. $f_3$) along its end-to-end path are $conv(f_2,f_1,7)=\{4,11\}$ (resp. $conv(f_3,f_1,7)=\{5\}$).
\end{itemize}

\subsection{Service Curve for a Flow of Interest}
\label{sec:GPMOC}
\begin{Corollary}
\label{th:GPMOC}(Service Curve under Arbitrary Multiplexing)
The service curve offered to a $f.o.i$ $f$ along its subpath, $\mathbb{P}_f (n)$, in a multiple-ring network under arbitrary multiplexing with strict service curve nodes of the rate-latency form $\beta_{R,T}$ and leaky bucket constrained arrival curves $\gamma_{\sigma, \rho}$, is a rate-latency curve, with a rate $R^{\mathbb{P}_f (n)}$ and a latency $T^{\mathbb{P}_f (n)}$,  defined as follows:
%========================================================================================================================================================
\begin{subequations}
\label{PMOC-service}
\begin{align}
&  R^{\mathbb{P}_f(n)} = \min \limits_{k \in \mathbb{P}_f(n)} [ R^{k}- \sum \limits_{i \ni k, i \neq f }{\rho_i} ] \label{PMOC-serviceR}\\
& T^{\mathbb{P}_f(n)} =  \sum\limits_{k \in \mathbb{P}_f(n)}T^{k} + \sum \limits_{i \in \mathbb{K}_f(n)} \frac{\sum\limits_{k\in conv(i,f,n)}{\sigma_i^{k\ominus 1}}+ \rho_i. \sum\limits_{j \in \mathbb{P}_f(n)\cap \mathbb{P}_i} T^{j}}{R^{\mathbb{P}_f(n)}} \label{PMOC-serviceT}
\end{align}
\end{subequations}
\end{Corollary}

As shown in Eq. (\ref{PMOC-serviceT}), some flow bursts are payed several times. These particular flows have actually more than one convergence point with the \textit{f.o.i}; thus respecting the principle of the PMOC approach.The proof of Cor. \ref{th:GPMOC} is provided in appendix \ref{proof2}. 

\textit{We detail here the end-to-end service curve of the \textit{f.o.i} $f_1$ in the example of Fig. \ref{fig:exampleM}, when the assumptions of the system model detailed in Sections \ref{Model} and \ref{Model-ext} are fulfilled, and all the crossed nodes offer the same service curve $\beta_{R,T}$. According to Cor. \ref{th:GPMOC}, this service curve is a rate-latency curve, with a rate $R^{\mathbb{P}_{f_1}(7)}=\min[R-\rho_2,R-\rho_3]$ and a latency $T^{\mathbb{P}_{f_1}(7)}=7.T+\frac{1}{R^{\mathbb{P}_{f_1}(7)}}.( \sigma_2^0 +\sigma_2^9 +2.\rho_2.T + \sigma_3^6 + \rho_3.T)$.} \\

Afterwards, we extend such a result to the FP multiplexing case, based on the same notations presented in Section \ref{sec:service}.

\begin{Corollary}(Service Curve under FP Multiplexing)
\label{cor:PMOC-FP}
The service curve offered to a $f.o.i$ $f$ along its subpath, $\mathbb{P}_f(n)$, in a multiple-ring network under FP multiplexing with strict service curve nodes of the rate-latency type $\beta_{R,T}$ and leaky bucket constrained arrival curves $\gamma_{\sigma, \rho}$, is a rate-latency curve, with a rate $R^{\mathbb{P}_f (n)}$ and a latency $T^{\mathbb{P}_f (n)}$,  defined as follows:
%========================================================================================================================================================
\begin{equation}
\label{PMOC-service-FP}
\begin{array}{lll}
R^{\mathbb{P}_f(n)} = \min \limits_{k \in \mathbb{P}_f(n)} [ R^{k}- \sum \limits_{i \ni hp_f^k}{\rho_i} ] \\
T^{\mathbb{P}_f(n)}  = \sum\limits_{k \in \mathbb{P}_f(n)}(T^{k} + \frac{\max_{i \in lp_f^k} L_{max}(i)}{R^k}) \\
+  \sum\limits_{i \in \mathbb{K}_{\leq f}(n)} \frac{\sum\limits_{k\in conv(i,f,n)}{\sigma_i^{k\ominus1}}+ \rho_i \cdot \sum\limits_{k \in \mathbb{P}_f(n)\cap \mathbb{P}_i} (T^{k}+ \frac{\max_{j \in lp_f^k} L_{max}(j)}{R^k})} {R^{\mathbb{P}_f(n)}}
\end{array} 
\end{equation}
\end{Corollary}

\begin{proof}
The proof of Cor. \ref{cor:PMOC-FP} is based on the same idea than Cor. \ref{Th:PMOO-Cycle-FP}.
\end{proof}

It is worth noting that the second step of the PMOC approach, which consists in computing the delay bound, remains the same as explained in Sec. \ref{sec:computation} under Arbitrary and FP multiplexing. First, we need to express the $\mathbb{M^*}$ parameters, then to verify the necessary and sufficient conditions defined in Corollaries \ref{Th:EUD} and \ref{Th:EUD-FP}.% ***************************************** ***************************************** ************************

\textbf{Example}\\
We now explicit the matrix form $\mathbb{M}^*$ and the necessary and sufficient condition on the existence of delay bounds for the example in Fig. \ref{fig:exampleM}. We consider that each flow $f_i$ is $(\sigma_i^0,\rho_i)$-constrained and each node $i$ guarantees a service curve $\beta_{R,T_i}$ under arbitrary multiplexing.

First, we express the formulas of the different parameters of the matrix form $\mathbb{M^*}$: ($A1, A2, C1, C2, T$).
\[
T^T=\scriptsize
\begin{array}{c}
(T^{\mathbb{P}_{f_1(1)}},
T^{\mathbb{P}_{f_1(2)}}
T^{\mathbb{P}_{f_1(3)}},
T^{\mathbb{P}_{f_1(4)}},
T^{\mathbb{P}_{f_1(5)}},
T^{\mathbb{P}_{f_1(6)}},
T^{\mathbb{P}_{f_1(7)}},
T^{\mathbb{P}_{f_2(1)}},
T^{\mathbb{P}_{f_2(2)}},
T^{\mathbb{P}_{f_2(3)}},\\
T^{\mathbb{P}_{f_2(4)}},
T^{\mathbb{P}_{f_2(5)}},
T^{\mathbb{P}_{f_2(6)}},
T^{\mathbb{P}_{f_2(7)}},
T^{\mathbb{P}_{f_3(1)}},
T^{\mathbb{P}_{f_3(2)}},
T^{\mathbb{P}_{f_3(3)}},
T^{\mathbb{P}_{f_3(4)}},
T^{\mathbb{P}_{f_3(5)}},
T^{\mathbb{P}_{f_3(6)}})
\end{array}\]

\[
C1=\left(\scriptsize
\begin{array}{c}
T_1\\
T_1+T_2\\
T_1+T_2+T_4+\frac{\rho_2 T_4}{R-\rho_2}\\
T_1+T_2+T_4+T_5+\frac{\rho_2 T_4+\rho_3 T_5}{R-\max(\rho_2,\rho_3)}\\
T_1+T_2+T_4+T_5+T_9+\frac{\rho_2 T_4+\rho_3 T_5}{R-\max(\rho_2,\rho_3)}\\
T_1+T_2+T_4+T_5+T_9+T_{10}+\frac{\rho_2 T_4+\rho_3 T_5}{R-\max(\rho_2,\rho_3)}\\
T_1+T_2+T_4+T_5+T_9+T_{10}+T_{11}+\frac{\sigma_2^0+\rho_2 (T_4+T_{11})+\rho_3 T_5}{R-\max(\rho_2,\rho_3)}\\
T_{11}+\frac{\rho_1 T_{11}}{R-\rho_1}\\
T_{11}+T_{12}+\frac{\rho_1 T_{11}}{R-\rho_1}\\
T_{11}+T_{12}+T_{10}+\frac{\rho_1 T_{11}}{R-\rho_1}\\
T_{11}+T_{12}+T_{10}+T_{9}+\frac{\rho_1 T_{11}+\rho_3 T_{9}}{R-\max(\rho_1,\rho_3)}\\
T_{11}+T_{12}+T_{10}+T_{9}+T_{4}+T_{5}+T_{6}+\frac{\rho_1 (T_{11}+T_{4})+\rho_3 T_9}{R-\max(\rho_1,\rho_3)}\\
T_{11}+T_{12}+T_{10}+T_{9}+T_{4}+T_{5}+T_{6}+\frac{\rho_1 (T_{11}+T_{4})+\rho_3 T_9}{R-\max(\rho_1,\rho_3)}\\
T_{11}+T_{12}+T_{10}+T_{9}+T_{4}+T_{5}+T_{6}+\frac{\rho_1 (T_{11}+T_{4})+\rho_3 T_9}{R-\max(\rho_1,\rho_3)}\\
T_8\\
T_8+T_6\\
T_8+T_6+T_5+\frac{\rho_1 T_5}{R-\rho_1}\\
T_8+T_6+T_5+T_9+\frac{\rho_1 T_5+\rho_2 T_9}{R-\max(\rho_1,\rho_2)}\\
T_8+T_6+T_5+T_9+T_4+\frac{\rho_1 T_5+\rho_2 T_9}{R-\max(\rho_1,\rho_2)}\\
T_8+T_6+T_5+T_9+T_4+T_2+\frac{\rho_1 T_5+\rho_2 T_9}{R-\max(\rho_1,\rho_2)} 
\end{array}\right)
\]

\[\sigma^T=\scriptsize
\begin{array}{c}
(\sigma_1^1,
\sigma_1^2,
\sigma_1^4,
\sigma_1^5,
\sigma_1^9,
\sigma_1^{10},
\sigma_1^{11},
\sigma_2^{11},
\sigma_2^{12},
\sigma_2^{10},\\
\sigma_2^9,
\sigma_2^4,
\sigma_2^5,
\sigma_2^6,
\sigma_3^8,
\sigma_3^6,
\sigma_3^5,
\sigma_3^9,
\sigma_3^4,
\sigma_3^2)
\end{array}\]

\[C2^T=\scriptsize
\begin{array}{c}
(\sigma_1^0,
\sigma_1^0,
\sigma_1^0,
\sigma_1^0,
\sigma_1^0,
\sigma_1^0,
\sigma_1^0,
\sigma_2^0,
\sigma_2^0,
\sigma_2^0,\\
\sigma_2^0,
\sigma_2^0,
\sigma_2^0,
\sigma_2^0,
\sigma_3^0,
\sigma_3^0,
\sigma_3^0,
\sigma_3^0,
\sigma_3^0,
\sigma_3^0)
\end{array}\]

\[\scriptsize{A2}=
\left(\scriptsize
\begin{array}{cccccccccccccccccccc}
\rho_1&0&0&0&0&0&0&0&0&0&0&0&0&0&0&0&0&0&0&0 \\
0&\rho_1&0&0&0&0&0&0&0&0&0&0&0&0&0&0&0&0&0&0 \\
0&0&\rho_1&0&0&0&0&0&0&0&0&0&0&0&0&0&0&0&0&0 \\
0&0&0&\rho_1&0&0&0&0&0&0&0&0&0&0&0&0&0&0&0&0 \\
0&0&0&0&\rho_1&0&0&0&0&0&0&0&0&0&0&0&0&0&0&0 \\
0&0&0&0&0&\rho_1&0&0&0&0&0&0&0&0&0&0&0&0&0&0 \\
0&0&0&0&0&0&\rho_1&0&0&0&0&0&0&0&0&0&0&0&0&0 \\
0&0&0&0&0&0&0&\rho_2&0&0&0&0&0&0&0&0&0&0&0&0 \\
0&0&0&0&0&0&0&0&\rho_2&0&0&0&0&0&0&0&0&0&0&0 \\
0&0&0&0&0&0&0&0&0&\rho_2&0&0&0&0&0&0&0&0&0&0 \\
0&0&0&0&0&0&0&0&0&0&\rho_2&0&0&0&0&0&0&0&0&0 \\
0&0&0&0&0&0&0&0&0&0&0&\rho_2&0&0&0&0&0&0&0&0 \\
0&0&0&0&0&0&0&0&0&0&0&0&\rho_2&0&0&0&0&0&0&0 \\
0&0&0&0&0&0&0&0&0&0&0&0&0&\rho_2&0&0&0&0&0&0 \\
0&0&0&0&0&0&0&0&0&0&0&0&0&0&\rho_3&0&0&0&0&0 \\
0&0&0&0&0&0&0&0&0&0&0&0&0&0&0&\rho_3&0&0&0&0 \\
0&0&0&0&0&0&0&0&0&0&0&0&0&0&0&0&\rho_3&0&0&0 \\
0&0&0&0&0&0&0&0&0&0&0&0&0&0&0&0&0&\rho_3&0&0 \\
0&0&0&0&0&0&0&0&0&0&0&0&0&0&0&0&0&0&\rho_3&0 \\
0&0&0&0&0&0&0&0&0&0&0&0&0&0&0&0&0&0&0&\rho_3    
\end{array}\right)
\]

%\begin{landscape}
{
\hspace{-2em}
\[A1=
\left(\scriptsize
\begin{array}{cccccccccccccccccccc}
0&0&0&0&0&0&0&0&0&0&0&0&0&0&0&0&0&0&0&0 \\
0&0&0&0&0&0&0&0&0&0&0&0&0&0&0&0&0&0&0&0 \\
0&0&0&0&0&0&0&0&0&0&\frac{1}{R-\rho_2}&0&0&0&0&0&0&0&0&0 \\
0&0&0&0&0&0&0&0&0&0&\frac{1}{R-\max(\rho_2,\rho_3)}&0&0&0&0&0&\frac{1}{R-\max(\rho_2,\rho_3)}&0&0&0 \\
0&0&0&0&0&0&0&0&0&0&\frac{1}{R-\max(\rho_2,\rho_3)}&0&0&0&0&0&\frac{1}{R-\max(\rho_2,\rho_3)}&0&0&0 \\
0&0&0&0&0&0&0&0&0&0&\frac{1}{R-\max(\rho_2,\rho_3)}&0&0&0&0&0&\frac{1}{R-\max(\rho_2,\rho_3)}&0&0 &0\\
0&0&0&0&0&0&0&0&0&0&\frac{1}{R-\max(\rho_2,\rho_3)}&0&0&0&0&0&\frac{1}{R-\max(\rho_2,\rho_3)}&0&0 &0\\
0&0&0&0&0&\frac{1}{R-\rho_1}&0&0&0&0&0&0&0&0&0&0&0&0&0&0 \\
0&0&0&0&0&\frac{1}{R-\rho_1}&0&0&0&0&0&0&0&0&0&0&0&0&0&0 \\
0&0&0&0&0&\frac{1}{R-\rho_1}&0&0&0&0&0&0&0&0&0&0&0&0&0&0 \\
0&0&0&0&0&\frac{1}{R-\max(\rho_1,\rho_3)}&0&0&0&0&0&0&0&0&0&0&\frac{1}{R-\max(\rho_1,\rho_3)}&0&0&0 \\
0&\frac{1}{R-\max(\rho_1,\rho_3)}&0&0&0&\frac{1}{R-\max(\rho_1,\rho_3)}&0&0&0&0&0&0&0&0&0&0&\frac{1}{R-\max(\rho_1,\rho_3)}&0&0&0 \\
0&\frac{1}{R-\max(\rho_1,\rho_3)}&0&0&0&\frac{1}{R-\max(\rho_1,\rho_3)}&0&0&0&0&0&0&0&0&0&0&\frac{1}{R-\max(\rho_1,\rho_3)}&0&0&0 \\
0&\frac{1}{R-\max(\rho_1,\rho_3)}&0&0&0&\frac{1}{R-\max(\rho_1,\rho_3)}&0&0&0&0&0&0&0&0&0&0&\frac{1}{R-\max(\rho_1,\rho_3)}&0&0&0 \\
0&0&0&0&0&0&0&0&0&0&0&0&0&0&0&0&0&0&0&0 \\
0&0&0&0&0&0&0&0&0&0&0&0&0&0&0&0&0&0&0&0 \\
0&0&\frac{1}{R-\rho_1}&0&0&0&0&0&0&0&0&0&0&0&0&0&0&0&0&0 \\
0&0&\frac{1}{R-\max(\rho_1,\rho_2)}&0&0&0&0&0&0&\frac{1}{R-\max(\rho_1,\rho_2)}&0&0&0&0&0&0&0&0&0&0 \\
0&0&\frac{1}{R-\max(\rho_1,\rho_2)}&0&0&0&0&0&0&\frac{1}{R-\max(\rho_1,\rho_2)}&0&0&0&0&0&0&0&0&0&0 \\
0&0&\frac{1}{R-\max(\rho_1,\rho_2)}&0&0&0&0&0&0&\frac{1}{R-\max(\rho_1,\rho_2)}&0&0&0&0&0&0&0&0&0&0   
\end{array}\right)
\]
}
%\end{landscape}

Second, to verify the necessary and sufficient condition defined in Cor. \ref{Th:EUD} and the symbolic computation of the determinant of the matrix $(Id-A_1\times A_2)$, we consider that all flows are $(\sigma^0,\rho)$-constrained and each node guarantees a service curve $\beta_{R,0}$. Hence, the determinant of the matrix $(Id-A_1\times A_2)$ becomes as follows:

\[-R(3\rho^3 -5R\rho^2+4R^2\rho -R^3)/(R-\rho)^{4}\]

This function vanishes for the flow rate $\rho=\frac{45-11R}{81(\sqrt{\frac{31}{8748}}-\frac{47}{1458})^\frac{1}{3}}$. Hence, if the flow rate condition is verified, i.e., $\rho < \frac{45-11R}{81(\sqrt{\frac{31}{8748}}-\frac{47}{1458})^\frac{1}{3}}$, then the end-to-end delay upper bound of the \textit{f.o.i} $f_1$, $EED_{f_1}^{\mathbb{P}_{f_1}(7)}$, exists and is at most equal to $\frac{\sigma^0 }{R^{\mathbb{P}_{f_1}(7)}} + T^{\mathbb{P}_{f_1}(7)}$, where $R^{\mathbb{P}_{f_1}(7)}=R-\rho$ and $T^{\mathbb{P}_{f_1}(7)} = \frac{\sigma^0}{R-\rho}(3 +\frac{-\rho(4\rho^2-5R\rho+3R^2)}{3\rho^3-5R\rho^2 + 4R^2\rho - R^3})$.

\subsection{Performance Evaluation}
\label{sec:eval-multi}
In this section, we investigate the offered timing performance of a multiple-ring topology, with respect to the inter-ring communication load \textit{interNet} and the number of rings \textit{nbR} to show their impact on the performance.

Hence, we compute the end-to-end delay bounds under different configurations according to the set of parameters $(interNet, M, nbR, \sigma, \rho)$.

We consider the case study with the following assumptions:
\begin{itemize}
	\item The network is based on a mono or multiple-ring topology with $nbR$ rings, connecting $M$ nodes, i.e., each ring connects $\frac{M-nbR}{nbR}$ nodes;
	\item Each node guarantees a service curve $\beta_{R=1Gbps, T=600ns}$;
	\item Each node generates one leaky-bucket constrained flow with a burst $\sigma$ and a rate $\rho$.
	\item Communications within each peripheral ring are broadcast.
\end{itemize}
	
	%================================================================ 
\begin{figure}[htbp]
	\centering
		\includegraphics[width=0.6\textwidth]{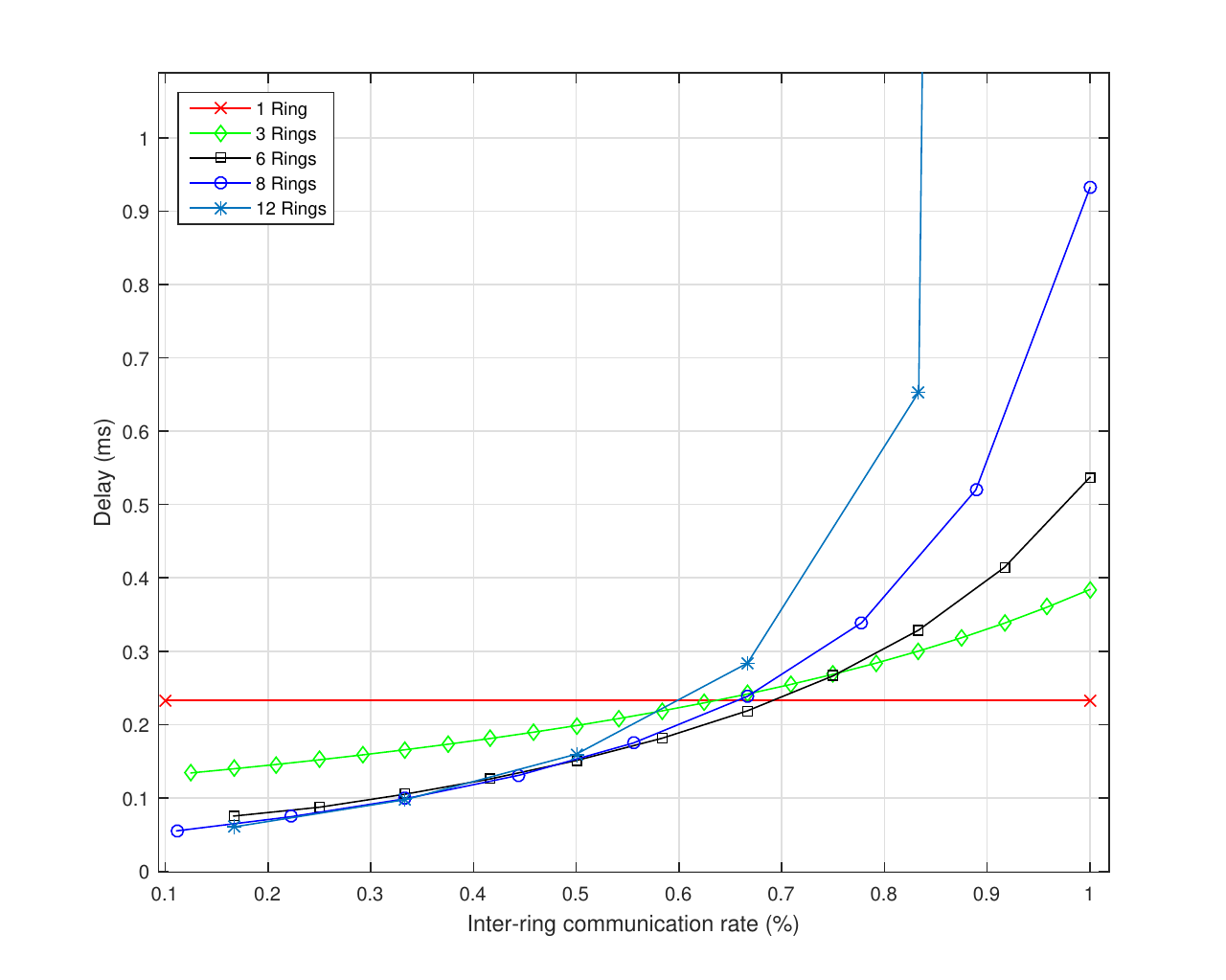}
	\caption{The impact of number of rings on delay bounds vs inter-ring communication load for $(interNet\in[0.2 - 1], M=72, \sigma=128bytes, \rho=5\cdot10^5bit/s)$.}
	\label{fig:delayInter}
\end{figure}
%========================================================

Fig. \ref{fig:delayInter} shows the impact of the inter-ring communication load and the number of rings on the end-to-end delay bounds. As we can see, the multiple-ring network is more sensitive to the inter-ring communication load when the number of rings increases, e.g., the 12-rings network offers the best delay bounds for an inter-ring communication load less than $34.8\%$, whereas, it guarantees the highest delay bounds for a load higher than $59\%$.
This behavior is due to the following facts:
\begin{enumerate}
	\item First, it is worth noting that the number of convergence points increases with the number of rings. Hence, the more this parameter increases, the more the delay bounds may increase;
	\item Second, increasing the inter-ring communication load leads to a higher impact of interfering flow at each convergence point.
\end{enumerate}

%====== 
\begin{figure}[htbp]
	\centering
		\includegraphics[width=0.6\textwidth]{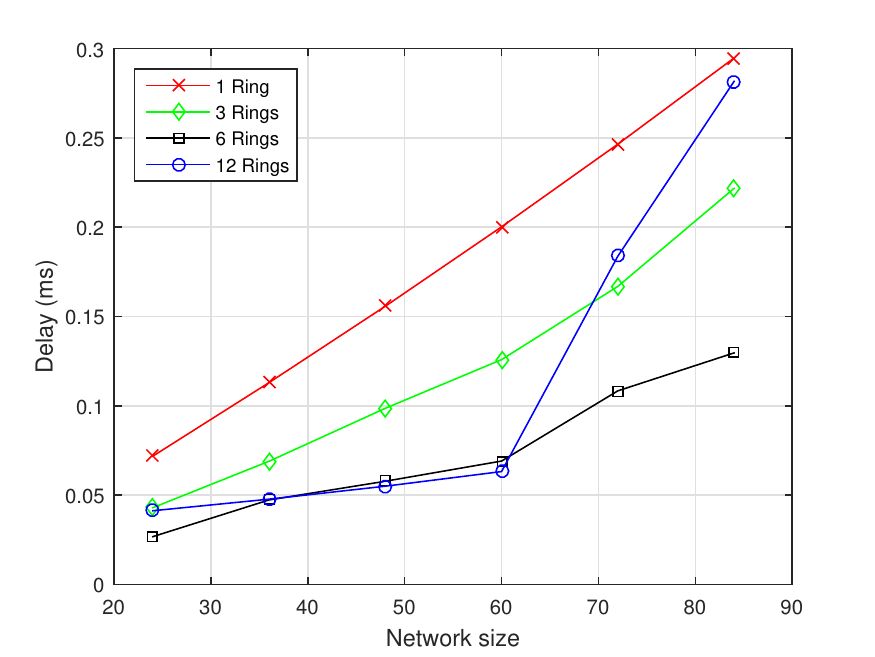}
	\caption{The impact of number of rings on delay bounds vs the network size, $(interNet=0.2, M=[24-84], \sigma=128bytes, \rho=10^6bit/s)$.}
	\label{fig:delayNodeInter}
\end{figure}

Fig. \ref{fig:delayNodeInter} shows the impact of the network size and the number of rings on the end-to-end delay bounds. As it is shown, the delay bounds are generally decreasing when increasing the number of rings. This is mainly due to the decreasing flow path length. In the worst-case for the multiple-ring case, a flow needs to cross the source peripheral ring, the backbone ring and the destination peripheral ring to reach its destination. Hence, the path length is equal to $M_{cross}=\frac{2M-nbR}{nbR}+nbR-1<M$ when $nbR>2$. Moreover, the delay bounds for the 12-rings topology increases dramatically for a network size higher than 60 nodes. This is mainly related to the increasing inter-ring communication load due to the increasing network size, which leads to a higher impact of interfering flows at each convergence point, as illustrated in Fig. \ref{fig:delayInter}.

\begin{figure}[htbp]
	\centering
		\includegraphics[width=0.6\textwidth]{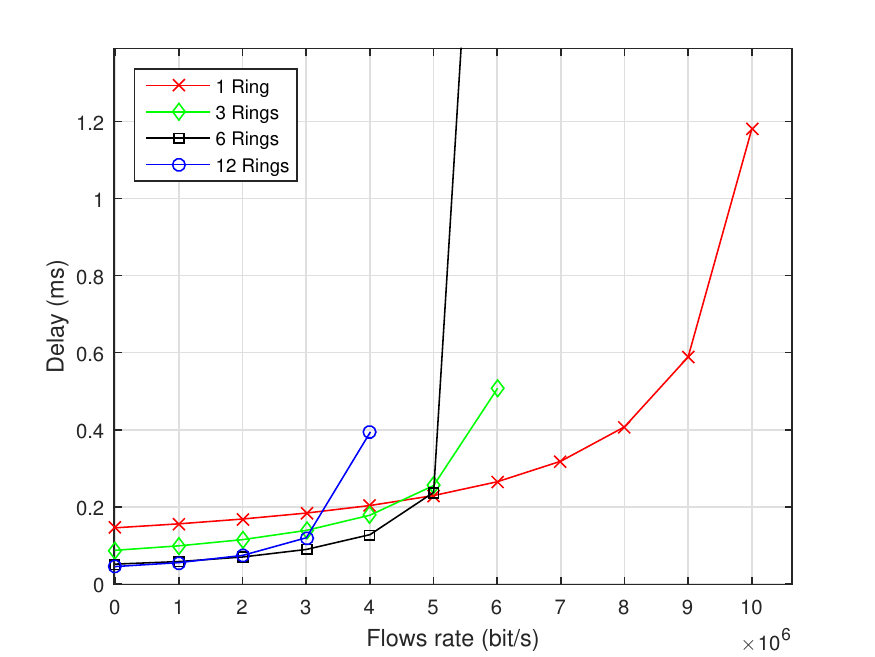}
	\caption{The impact of number of rings on delay bounds vs the flows rate, $(interNet=0.2, M=48, \sigma=128bytes, \rho=[10^3-10^7]bit/s)$.}
	\label{fig:delayRateInter}
\end{figure}

Fig. \ref{fig:delayRateInter} shows the impact of the flows rate on the end-to-end delay bounds. We observe that the multiple-ring network is more sensitive to the flows rate when the number of rings increases, i.e., the 12-rings network offers the lowest delay bounds for a rate up to $10^3$bit/s, however it is the first to lead to the delay bound divergence for a rate higher than $4\times10^6$ bit/s. 

\begin{figure}[htbp]
	\centering
		\includegraphics[width=0.6\textwidth]{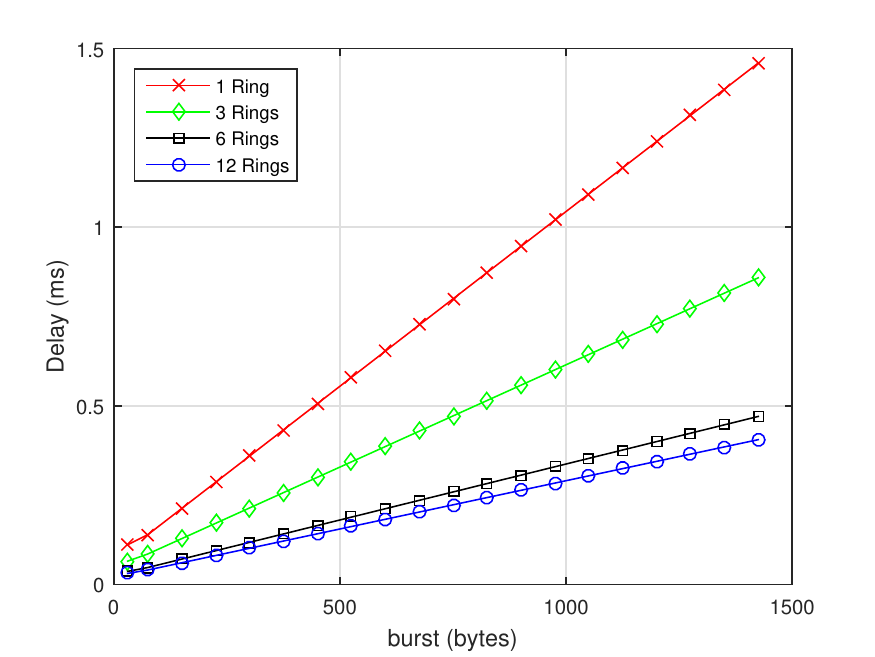}
	\caption{The impact of number of rings on delay bounds vs the flows burst, $(interNet=0.2, M=60, \sigma=[30-1500] bytes, \rho=5\times10^5 bit/s)$.}
	\label{fig:delayBurstInter}
\end{figure}

This fact is mainly due to the violation of the network utilization rate condition in Cor. \ref{Th:EUD}. On the other hand, we can also observe from Fig. \ref{fig:delayBurstInter} that multiple-ring topology is less sensitive to the flows burst than flow rate, i.e., the more the number of rings increases, the more the delay bounds decreases.

%====== 
%========================================================

%====== 
%========================================================

\textbf{These results have shown that the end-to-end delay bounds of the multiple-ring topology are particularly sensitive to the inter-ring communication load and the flows rate. For a low inter-ring communication load, dividing the network into several rings may improve the end-to-end delay bounds, since it reduces the impact of interfering flows and the path length. However, for a high inter-ring communication load, the impact of convergence points increases with the number of rings, which leads to increasing the delay bounds.}

%###################################
\section{Avionics Case Study}
\label{UseCase}
\begin{figure}[htbp]
	\centering
		\includegraphics[width=0.5\textwidth]{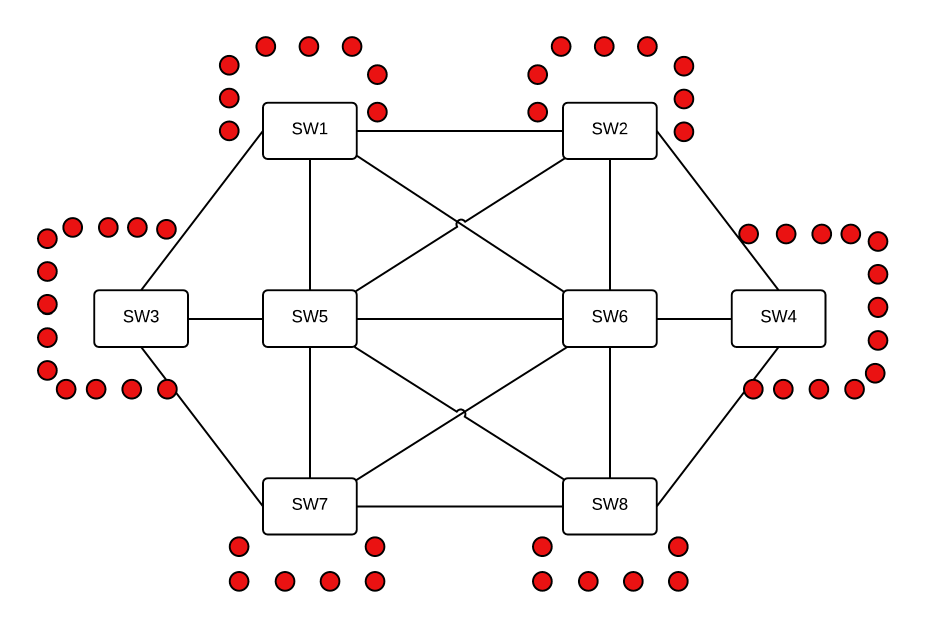}
	\caption{A representative A380 AFDX network}
	\label{fig:AFDXcaseStudy}
\end{figure}

%###################################
In this section, we will illustrate the usage of our proposed approach PMOC to analyse the performance of realistic multiple-ring networks. The considered case study is a representative avionics backbone network of an A380. As shown in Fig. \ref{fig:AFDXcaseStudy}, it consists of 8 AFDX switches \cite{afdx} connecting 54 end-systems, where, each end-system sends 8 traffic flows from 3 different traffic classes as described in Tab. \ref{tab:TC}.

%================================================================ 
%========================================================

%================================================================
\begin{table}[!h]
  \centering
  \caption{Traffic Classes}
    \begin{tabular}{|c|c|c|c|c|}
    \hline
		TC & Period (ms) & Payload size (byte) & Rate (bit/s) & \# Flows/end-system\\
    \hline\hline
    1 & 4 & 480 &  $1024\times10^3$ & 1\\
    \hline
    2 & 8  & 16 &  $72\times10^3$ & 1  \\
    \hline
    3 & 32 & 480 &  $128\times10^3$ & 6\\
    \hline
    \end{tabular}
  \label{tab:TC}
\end{table}
%=========================================================================================

\begin{figure}[!htb]
    \centering
    	\subfigure[6-Rings]{
    	\label{Fig:6rings}
        \includegraphics[width=0.3\textwidth]{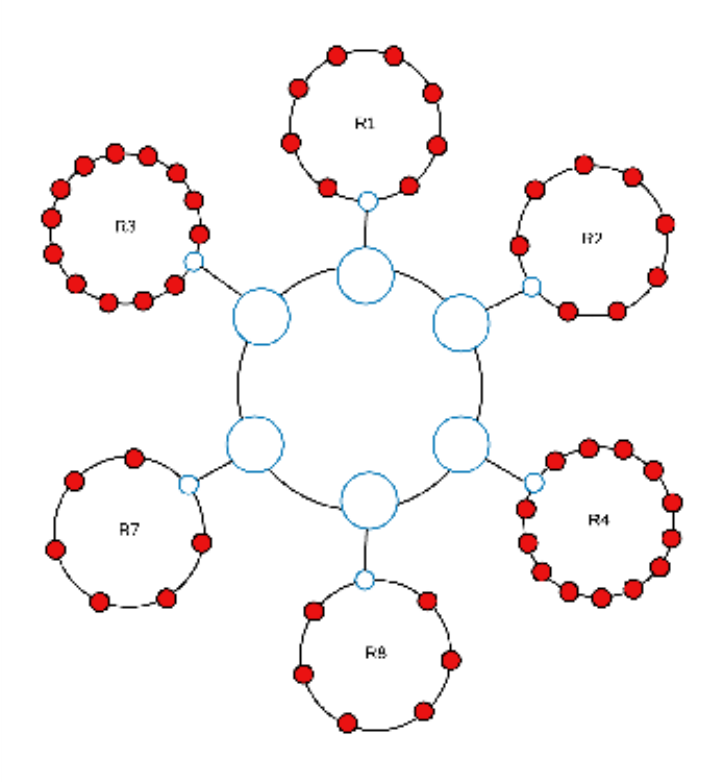}
        }
       \subfigure[4-Rings]{
        \label{fig:4rings}
        \includegraphics[width=0.3\textwidth]{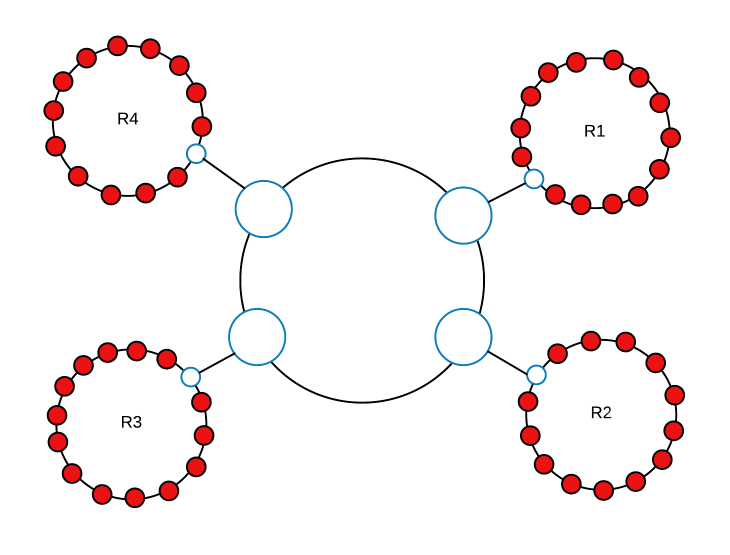}
       }
  	\subfigure[3-Rings]{
   	\label{fig:3rings}
        \includegraphics[width=0.3\textwidth]{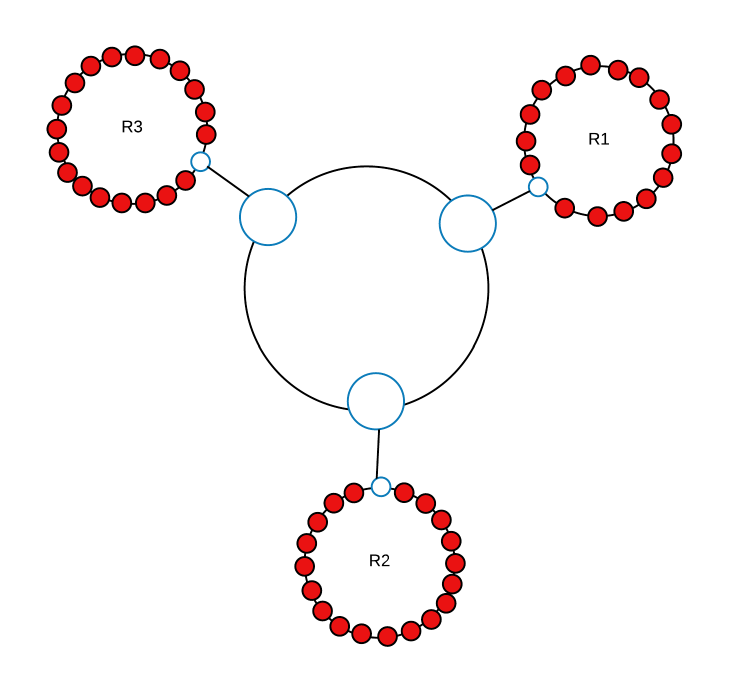}
      }
    \caption{Considered Multiple-Ring Topologies for the Avionics Case Study}
\end{figure}

The aim is to assess the performance of different multiple-ring configurations based on AeroRing technology \cite{amari2017theses,amari2016aeroring} in terms of delay bounds, with reference to the AFDX network \cite{afdx}. The considered multiple-ring topologies are: 
\begin{itemize}
	\item The 6-rings topology, described in Fig. \ref{Fig:6rings} and Table \ref{tab:multiRConf}, where we replace each AFDX switch by a peripheral ring;
	\item The 4-rings topology as described in Fig. \ref{fig:4rings} and Table \ref{tab:multiRConf}, where switches SW3 and SW4 are replaced each by a peripheral ring, whereas switches SW1 and SW7 (resp. SW2 and SW8) are grouped within the same peripheral ring;
	\item The 3-rings topology as described in Fig. \ref{fig:3rings} and Table \ref{tab:multiRConf}, where each couple of switches among (SW1, SW2), (SW3, SW7) and (SW4, SW8) is replaced by one peripheral ring;
	\item The mono-ring topology, where all the end-systems of the AFDX switches are gathered in the same ring.
\end{itemize}

The considered service policy within the nodes for all configurations is FP and the delay bounds under AFDX are computed according to \cite{Grieu04} and using WoPANets tool \cite{Wopanets-url}.

%================================================================
\begin{table}[!h]
  \centering
  \caption{Multiple-ring configurations}
    \begin{tabular}{|c|c|c|c|c|}
    \hline
		Peripheral ring id & 6 rings & 4 rings & 3 rings & 1 ring\\
    \hline\hline
    R1 & SW1 & SW1+SW7 &  SW1+SW2 & SW1+SW2+SW3\\
		   &     &         &          &+SW4+SW7+SW8 \\
    \hline
    R2 & SW2  & SW3 &  SW4+SW8 & - \\
    \hline
    R3 & SW3 & SW2+SW8 &  SW3+SW7 & - \\
    \hline
    R4 & SW4 & SW4 &  - & - \\
    \hline
    R5 & SW7  & - &  - & - \\
    \hline
    R6 & SW8 & - &  - & - \\
    \hline
    \end{tabular}
  \label{tab:multiRConf}
\end{table}

%================================================================ 
\begin{figure}[htbp]
	\centering
		\includegraphics[width=0.8\textwidth]{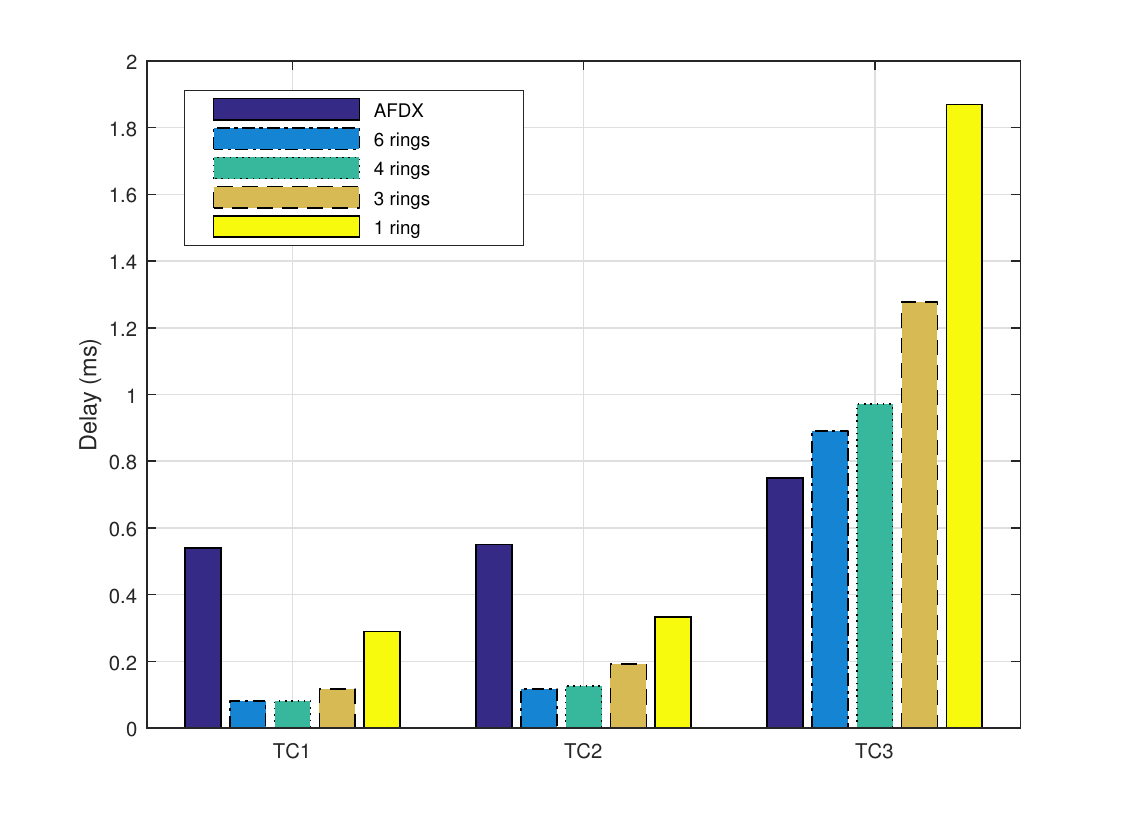}
	\caption{Maximum end-to-end delay bounds per traffic class}
	\label{fig:useCase}
\end{figure}
%========================================================

Fig. \ref{fig:useCase} shows the end-to-end delay bounds of the different traffic classes described in Table \ref{tab:TC}, when considering the current AFDX network and the different multiple-ring configurations. As we can notice, all the network solutions respect the temporal constraints of the different traffic classes, i.e., periods. Moreover, the multiple-ring networks outperform the AFDX in terms of TC1 and TC2 delay bounds; whereas, they offer a slightly higher delay bound for TC3. For instance, the 6-rings network offers a delay bound for TC1 $4.57$ times lower than the AFDX one. These results show the high timing performance of such an architecture with reference to AFDX. Moreover, we notice that the end-to-end delay bounds increase when reducing the number of peripheral rings, i.e., increasing the peripheral rings size. This fact is due to the increasing number of crossed nodes when the peripheral ring size increases, i.e., increasing the flows path length, which increases the interferences. These results are consistent with the conclusion of the performance evaluation of multiple-ring networks in Section \ref{sec:eval-multi}, since the inter-ring communication load is low for this avionics case of study.

%###################################%###################################
\section{Conclusion}
\label{Conclusions}
%###################################%###################################
In this paper, we have introduced a new approach, called PMOC, to compute delay bounds in multiple-ring networks with cyclic dependencies. This proposed approach integrates the flow serialization phenomena along the flow path, to allow the computation of tighter end-to-end delay bounds, with reference to existing timing analyses in this area. Hence, we have defined and proved the guaranteed end-to-end service curves of any f.o.i crossing such a network under Arbitrary and Fixed Priority multiplexing, for mono-ring and multiple-ring networks. Moreover, the computation of the delay bounds have been presented in the general case, and illustrated for the specific case of regular ring networks. Detailed sensitivity and tightness analyses have highlighted the accuracy of our proposed approach, in comparison to conventional methods and with reference to an achievable worst-case delay. This fact yields enhanced network performance in terms of resource efficiency and network scalability. Finally, the efficiency of our proposal has been illustrated through a realistic avionics case study.

The next step of this work is to extend the PMOC approach to compute delay bounds in the general case of non-feedforward networks.

\section*{Acknowledgements}
The authors would like to thank Prof. Jens B. Schmitt from University of Kaiserslautern for the fruitful discussion and interesting comments, which helped to improve this paper.
%\endacknowledgements

\newpage
\appendix

\section*{Appendix}

%###################################
%###################################
\section{Proof of Th \ref{Th:PMOO-Cycle}}
\label{proof}
%###################################
%###################################

\begin{proof}

As explained in Section \ref{sec:PMOCbasics}, for any flow $i$ crossing the ring network, there are only two possible convergence points with a f.o.i $f$: $f.ft$ and $i.ft$. This fact infers three possible categories for an interfering flow $i$ with the f.o.i $f$: (i) \textit{category 1}: having only one convergence point with $f$, which is its first hop, i.e., $i.ft$; (ii) \textit{category 2} having only one convergence point with $f$, which is the first hop of $f$, i.e., $f.ft$; (iii) \textit{category 3} having two distinct convergence points with $f$, i.e., $i.ft$ and $f.ft$ if $i.ft\neq f.ft$.

\textit{We illustrate these three categories with the example of Fig. \ref{figure:cycle}. If we consider flow $f_1$ as the f.o.i, then flows $f_2$, $f_4$ and $f_3$ are in categories $1$, $2$ and $3$, respectively.}

%=======================================================================================
\begin{figure}[htbp]
\centering     %%% not \center
\includegraphics[width=75mm]{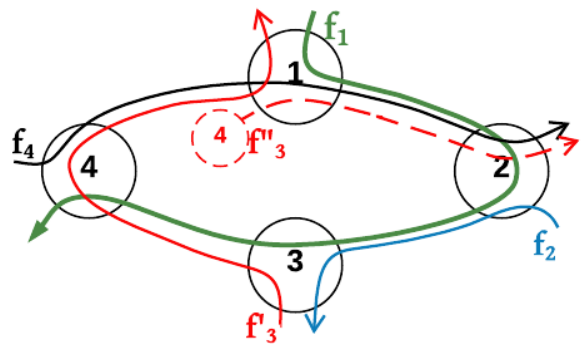}
\caption{Cutting virtually the flows of Fig. \ref{figure:cycle}}
\label{figure:Vcycle}
\end{figure}
%=====================================
To prove the Th. \ref{Th:PMOO-Cycle}, we need to model an interfering flow $i$ of category $3$ by splitting it in two subflows to cut virtually the cyclic dependency with the f.o.i $f$, as illustrated in Fig. \ref{figure:Vcycle} for flow $f_3$: (i) $i1$: the subflow of $i$ along its subpath $\mathbb{P}_{i1} = (0, i.ft,i.ft\oplus1, ...,f.ft\ominus1)$, which is $(\sigma_i^0, \rho_i)$-constrained; (ii) $i2$: the subflow of $i$ along its subpath $\mathbb{P}_{i2} = (f.ft\ominus1, f.ft, ..., i.ft\oplus(h_i- 1))$, which is $(\sigma_i^{f.ft\ominus 1}, \rho_i)$-constrained. It is worth noting that $i1$ fulfills the conditions of category $1$, whereas $i2$ fulfills the ones of category $2$. Thus, splitting virtually the flows of category $3$ in $\mathbb{K}_f(n)$ in two subflows leads to a transformed set $\overline {\mathbb{K}_f(n)}$. The latter can be rewritten according to the conditions of categories $1$ and $2$ as follows:
 \[\overline {\mathbb{K}_f(n)}= \{ i \in \overline{ \mathbb{K}_f(n)} / f \ni i.ft\} \cup \{ i \in \overline {\mathbb{K}_f(n)} / i \ni f.ft, i.ft \neq f.ft\}\]

\textit{We explicit $\overline {\mathbb{K}_f(n)}$ through the example of Fig. \ref{figure:cycle}. For the f.o.i $f_1$, the only flow of category $3$ is the flow $f_3$. So, $f_3$ is virtually splitted as $(f'_3, f"_3)$ as shown in Fig. \ref{figure:Vcycle}, where $\mathbb{P}_{f'_3} =\{0,3,4\}$ and $\mathbb{P}_{f"_3} =\{4, 1\}$. It is worth noting that according to this model, the virtual node representing the source of flow $f"_3$ is node $4$. Moreover, the set of interfering flows with the f.o.i $f_1$, $\mathbb{K}_{f_1}(3)$, is transformed to $\overline {\mathbb{K}_{f_1}(3)}= \{ f_2, f'_3 \} \cup \{ f_4, f"_3 \}$. }
 
Consider a flow of interest $f$ with a subpath $\mathbb{P}_f(n)$. Any crossed node $ l \in \mathbb{P}_f(n)$ admits a strict service curve. Hence, according to Def. \ref{def:strict-service-curve}, for any instant $t_l \geq 0$, there exists $t_{l\ominus 1} \leq t_l$ the start of the backlogged period such that:
\begin{equation}
\label{label1}
   D_f^{l}(t_l) - D_f^{l}(t_{l\ominus 1}) + \sum \limits_{i \ni l, i \neq f} ( D_i^{l}(t_{l})- D_i^{l}(t_{l\ominus 1}) ) \geq \beta^{l}(\Delta_l ) 
\end{equation}
where $\Delta_l = t_l - t_{l\ominus 1}$. The time indices are chosen to match the node indices. Then, we sum up the expression in Eq. (\ref{label1}) when varying $l \in \mathbb{P}_f(n)$, which infers:
\begin{eqnarray}
\label{label2}
&&  \sum  \limits_{ l \in \mathbb{P}_f(n)} D_f^{l}(t_l) - D_f^{l}(t_{l\ominus 1}) \\
&&  \geq \sum  \limits_{ l \in \mathbb{P}_f(n)} \beta^{l}(\Delta_l ) -  \sum  \limits_{ l \in \mathbb{P}_f(n)}  \sum \limits_{i \ni l, i \neq f} ( D_i^{l}(t_{l})- D_i^{l}(t_{l\ominus 1}))\nonumber
\end{eqnarray}

Knowing the definition of $\overline{\mathbb{K}_f(n)}$, we have:
 \[  \sum \limits_{ l \in \mathbb{P}_f(n)}  \sum  \limits_{i \ni l, i \neq f} \Leftrightarrow \sum  \limits_{i \in \overline{ \mathbb{K}_f(n)}} \sum  \limits_{l \in \mathbb{P}_f(n)\cap \mathbb{P}_i} \]
 
Moreover, at the start of a backlogged period $s$, we have $D_f^{i \oplus 1}(s)= A_f^{i \oplus 1}(s)$, and because of the ring topology, we have $A_f^{i \oplus 1}(s)= D_f^i(s)$; thus, $D_f^{i \oplus 1}(s) = D_f^i(s)$. Consequently, Eq . (\ref{label2}) can be simplified as follows:
\begin{eqnarray}
\label{label0}
&&  \sum  \limits_{ l \in \mathbb{P}_f(n)} D_f^{l}(t_l) - D_f^{l}(t_{l\ominus 1}) \\
&= &   D_f^{f.ft}(t_{f.ft}) - D_f^{f.ft}(t_{f.ft\ominus 1}) \nonumber\\
&+ &   D_f^{f.ft\oplus 1}(t_{f.ft\oplus 1}) - D_f^{f.ft \oplus 1}(t_{f.ft}) \nonumber\\
&...& \nonumber\\
&+ &   D_f^{f.ft\oplus (n-1)}(t_{f.ft\oplus (n-1)}) - D_f^{f.ft\oplus (n-1)}(t_{f.ft\oplus (n-2)}) \nonumber\\
&=&   D_f^{f.ft\oplus (n-1)}(t_{f.ft\oplus (n-1)}) - D_f^{f.ft}(t_{f.ft\ominus 1}) \nonumber \\
&\geq  &  \sum  \limits_{ l \in \mathbb{P}_f(n)} \beta^{l}(\Delta_l) -  \sum \limits_{i \in \overline{\mathbb{K}_f(n)}} \sum  \limits_{l \in \mathbb{P}_f(n)\cap \mathbb{P}_i}  (D_i^{l}(t_{l})- D_i^{l}(t_{l \ominus 1}) \nonumber
\end{eqnarray}

Based on the definitions of $ Mft(i, f, n)$ and $Mlt (i,f, n)$ in Tab. \ref{tab1}, Eq. (\ref{label0}) can be rewritten as follows:
\begin{eqnarray}
\label{label3}
&&  D_f^{f.ft\oplus (n-1)}(t_{f.ft\oplus (n-1)}) - D_f^{f.ft}(t_{f.ft\ominus 1} ) \\
&  \geq  & \sum \limits_{ l \in \mathbb{P}_f(n)} \beta^{l}(\Delta_l) -  \sum \limits_{i \in \overline{\mathbb{K}_f(n)}}  D_i^{Mlt (i,f, n)}(t_{Mlt (i,f, n)}) - D_i^{Mft(i, f, n)}(t_{Mft(i, f, n)\ominus 1}) \nonumber\\
&  \geq &  \sum \limits_{ l \in \mathbb{P}_f(n)} \beta^{l}(\Delta_l)  - \sum \limits_{i \in \overline{\mathbb{K}_f(n)}} A_i^{Mlt (i,f, n)}(t_{Mlt (i,f, n)}) - A_i^{Mft(i, f, n)}(t_{Mft(i, f, n)\ominus 1}) \nonumber\\
&\geq &  \sum \limits_{ l \in \mathbb{P}_f(n)} \beta^{l}(\Delta_l )  - \sum \limits_{i \in \overline{\mathbb{K}_f(n)}} \alpha_i^{Mft(i,f,n) \ominus 1}(\sum \limits_{l = Mft(i,f,n)}^{Mlt(i,f,n)} \Delta_l ) \nonumber 
\end{eqnarray}

To substitute the cumulative traffic functions of flows in $ \overline{\mathbb{K}_f(n)}$ in Eq. (\ref{label3}) by their arrival curves, we have used the causality constraint of cumulative traffic functions, i.e., $ \forall t, A_i^k(t) \geq D_i^k(t)$ and the property of the start of backlogged period at $t_{Mft(i, f, n)\ominus 1}$, i.e., $ D_i^{Mft(i, f, n)}(t_{Mft(i, f, n)\ominus 1})= A_i^{Mft(i, f, n)}(t_{Mft(i, f, n)\ominus 1})$. 

On the other hand, rewriting the input arrival curve of a flow $i$ at node $k$, $\alpha_i^{k\ominus 1}$, using $\overline{\alpha_i}(\Delta_l) = \rho_i \Delta_l$, infers:
\begin{eqnarray}
\label{label5}
 \alpha_i^{k \ominus 1}(\sum \limits_{l = 1}^{m} \Delta_l) & = &\sigma_i^{k \ominus 1} + \rho_i \sum \limits_{l = 1}^{m} \Delta_l \nonumber\\
 &= &\sigma_i^{k \ominus 1} + \rho_i \Delta_1+ \rho_i \sum \limits_{l = 2}^{m} \Delta_l \nonumber\\
&  = &\alpha_i^{k \ominus 1}(\Delta_1) +  \sum \limits_{l = 2}^{m} \overline{\alpha_i}(\Delta_l) 
\end{eqnarray}

Hence, Eq. (\ref{label3}) can be rewritten using Eq. (\ref{label5}) as follows:
\begin{eqnarray}
\label{label6}
&&  D_f^{f.ft\oplus (n-1)}(t_{f.ft\oplus (n-1)}) - D_f^{f.ft}(t_{f.ft\ominus 1}) \\
&\geq  &  \sum \limits_{ l \in \mathbb{P}_f(n)} [ \beta^{l}(\Delta_l ) - \sum \limits_{i \ni l, i \neq f} \alpha_i^{l \ominus 1}(\Delta_l). 1_{\{l=Mft(i,f,n) \} } + \overline{\alpha_i}(\Delta_l). 1_{\{l \neq Mft(i,f,n)\} }  ]  \nonumber \\
&\geq &  \sum  \limits_{ l \in \mathbb{P}_f(n)} [(R^l - \sum \limits_{i \ni l, i \neq f} \rho_i). (\Delta_l - T^l - \frac{\sum \limits_{i \ni l, i \neq f} \sigma_i^{Mft(i,f,n)\ominus 1} + T^l.\sum \limits_{i \ni l, i \neq f} \rho_i }{R^l - \sum \limits_{i \ni l, i \neq f} \rho_i})]^+  \nonumber\\
&\geq &  \min_{l \in \mathbb{P}_f(n)} (R^l - \sum \limits_{i \ni l, i \neq f} \rho_i). [\sum  \limits_{ l \in \mathbb{P}_f(n)} \Delta_l - \sum \limits_{ l \in \mathbb{P}_f(n)}T^l - \sum  \limits_{ l \in \mathbb{P}_f(n)} \frac{\sum \limits_{i \ni l, i \neq f} \sigma_i^{Mft(i,f,n)\ominus 1} + T^l.\sum \limits_{i \ni l, i \neq f} \rho_i }{R^l - \sum \limits_{i \ni l, i \neq f} \rho_i})  ]^+  \nonumber
\end{eqnarray}

Knowing the definition of $\overline {\mathbb{K}_f(n)}$, we can easily verify that
\[  \sum  \limits_{ l \in \mathbb{P}_f(n)} T^l.\sum \limits_{i \ni l, i \neq f} \rho_i \Leftrightarrow \sum \limits_{i \in \overline{\mathbb{K}_f(n)}} \rho_i. \sum\limits_{j \in \mathbb{P}_f(n)\cap \mathbb{P}_i} T^{j}\]

Hence, Eq. (\ref{label6}) becomes:
\begin{eqnarray}
\label{label8}
&&  D_f^{f.ft\oplus (n-1)}(t_{f.ft\oplus (n-1)}) - D_f^{f.ft}(t_{f.ft\ominus 1})  \\
& &  \geq \min_{l \in \mathbb{P}_f(n)} (R^l - \sum \limits_{i \ni l, i \neq f} \rho_i). [t_{f.ft\oplus (n-1)} - t_{f.ft\ominus 1} - \sum \limits_{ l \in \mathbb{P}_f(n)}T^l \nonumber\\
&&  - \sum \limits_{i \in \overline{\mathbb{K}_f(n)}} \frac{\sigma_i^{Mft(i,f,n)\ominus 1}+ \rho_i. \sum\limits_{j \in \mathbb{P}_f(n)\cap \mathbb{P}_i} T^{j}}{\min \limits_{l \in \mathbb{P}_f(n)} (R^l - \sum \limits_{i \ni l, i \neq f} \rho_i)} ] ^+\nonumber\\
& &  \geq \min_{l \in \mathbb{P}_f(n)} (R^l - \sum \limits_{i \ni l, i \neq f} \rho_i). [t_{f.ft\oplus (n-1)} - t_{f.ft\ominus 1} -  \sum \limits_{k \in \mathbb{P}_f(n)}T^{k} \nonumber\\
&& - \sum\limits_{i \in \overline {\mathbb{K}_f(n)}, f \ni i.ft} \frac{\sigma_i^{Mft(i,f,n)\ominus 1} + \rho_i \cdot \sum\limits_{j \in \mathbb{P}_f(n)\cap \mathbb{P}_i} T^{j}} {\min \limits_{l \in \mathbb{P}_f(n)} (R^l - \sum \limits_{i \ni l, i \neq f} \rho_i)} \nonumber\\
&& -  \sum\limits_{\underset{\underset{i.ft \neq f.ft}{i \ni f.ft}}{i \in \overline {\mathbb{K}_f(n)}}} \frac{ \sigma_i^{Mft(i,f,n)\ominus 1} + \rho_i \cdot \sum\limits_{j \in \mathbb{P}_f(n)\cap \mathbb{P}_i} T^{j}} {\min \limits_{l \in \mathbb{P}_f(n)} (R^l - \sum \limits_{i \ni l, i \neq f} \rho_i)} ]^+\nonumber
\end{eqnarray}

Moreover, for each interfering flow $i$ in category $3$ splitted as $(i_1, i_2)$, with $i_1$ and $i_2$ in categories $1$ and $2$, we have:
\begin{eqnarray}
\label{label9}
& &  \sigma_{i1}^{Mft(i_1,f,n)\ominus 1} + \rho_i. \sum\limits_{j \in \mathbb{P}_f(n)\cap \mathbb{P}_{i1}} T^{j} \\%\nonumber\\
&+&  \sigma_{i2}^{Mft(i_2,f,n)\ominus 1} + \rho_i. \sum\limits_{j \in \mathbb{P}_f(n)\cap \mathbb{P}_{i2}} T^{j} \nonumber\\
&= &  \sigma_{i}^{i.ft\ominus 1} + \sigma_i^{f.ft\ominus 1} + \rho_i. \sum\limits_{j \in \mathbb{P}_f(n)\cap (\mathbb{P}_{i1} \cup \mathbb{P}_{i2}) } T^{j} \nonumber\\
&= &  \sigma_i^{0} + \sigma_i^{f.ft\ominus 1} + \rho_i. \sum\limits_{j \in \mathbb{P}_f(n)\cap \mathbb{P}_i } T^{j} \nonumber
\end{eqnarray}
Using Eq. (\ref{label9}) and (\ref{label8}), we deduce:
\begin{eqnarray}
&&   R^{\mathbb{P}_f(n)} = \min_{l \in \mathbb{P}_f(n)} (R^l - \sum \limits_{i \ni l, i \neq f} \rho_i) \nonumber\\
&&  T^{\mathbb{P}_f(n)} = \sum\limits_{k \in \mathbb{P}_f(n)}T^{k} +  \sum\limits_{i \in \mathbb{K}_f(n)} \frac{  \sigma_i^{0}.1_{\{f \ni i.ft\}} + \rho_i \cdot \sum\limits_{j \in \mathbb{P}_f(n)\cap \mathbb{P}_i} T^{j}} {R^{\mathbb{P}_f(n)}} \nonumber\\
&& +  \sum\limits_{i \in \mathbb{K}_f(n)} \frac{  \sigma_i^{f.ft\ominus 1}.1_{\{i.ft \neq f.ft / i\ni f.ft\}}} {R^{\mathbb{P}_f(n)}} \\
\end{eqnarray}

This finishes the proof of the theorem.

\end{proof}

%###################################
%###################################
\section{Proof of Corollary \ref{th:GPMOC}}
\label{proof2}
%###################################
%###################################

\begin{proof}

In multiple-ring networks, an interfering flow $i$ can converge with the $f.o.i$ $f$ in several convergence points along its subpath of length $n$, denoted $conv(i,f,n)$. We need to model these flows by splitting them into several subflows, one subflow at each convergence point. Each subflow $i_k$, $k\in conv(i,f,n)$ has a path $\mathbb{P}_{i_k}$ and it is $(\sigma_{i_k}^0,\rho_i)-$constrained, where $Mft(i_k,f,n)=i_k.ft=k$ and $\sigma_{i_k}^0=\sigma_{i}^{k\ominus1}$. Thus, splitting the interfering flows in $\mathbb{K}_f(n)$ leads to a transformed set $\overline{ \mathbb{K}_f(n)}$.

We follow the same proof steps of Th. \ref{Th:PMOO-Cycle} from Eq. (\ref{label1}) to Eq. (\ref{label6}) in \ref{proof}. Then, knowing the definition of $\overline {\mathbb{K}_f(n)}$, we can easily verify that
\[  \sum  \limits_{ l \in \mathbb{P}_f(n)} T^l.\sum \limits_{i \ni l, i \neq f} \rho_i \Leftrightarrow \sum \limits_{i \in \overline{\mathbb{K}_f(n)}} \rho_i. \sum\limits_{j \in \mathbb{P}_f(n)\cap \mathbb{P}_i} T^{j}\]

Hence, Eq. (\ref{label6}) becomes:
\begin{eqnarray}
\label{label28}
&&  D_f^{f.ft\oplus (n-1)}(t_{f.ft\oplus (n-1)}) - D_f^{f.ft}(t_{f.ft\ominus 1})  \\
& &  \geq \min_{l \in \mathbb{P}_f(n)} (R^l - \sum \limits_{i \ni l, i \neq f} \rho_i). [t_{f.ft\oplus (n-1)} - t_{f.ft\ominus 1} - \sum \limits_{ l \in \mathbb{P}_f(n)}T^l \nonumber\\
&&  - \sum \limits_{i \in \overline{\mathbb{K}_f(n)}} \frac{\sigma_i^{Mft(i,f,n)\ominus 1}+ \rho_i. \sum\limits_{j \in \mathbb{P}_f(n)\cap \mathbb{P}_i} T^{j}}{\min \limits_{l \in \mathbb{P}_f(n)} (R^l - \sum \limits_{i \ni l, i \neq f} \rho_i)} ] ^+\nonumber
\end{eqnarray}

We have, $\sum\limits_{i\in \overline{\mathbb{K}_f(n)}} \sigma_{i}^{Mft(i,f,n)\ominus 1} = \sum\limits_{i\in \overline{\mathbb{K}_f(n)}} \sigma_{i}^{i.first\ominus 1} = \sum\limits_{i\in \mathbb{K}_f(n)} \sum\limits_{k \in conv(i,f,n)} \sigma_i^{k\ominus 1}$. Furthermore, the common shared path between the flow of interest $f$ and the original interfering flow $i$, i.e., $\mathbb{P}_f(n)\cap \mathbb{P}_i$, is equal to the shared path between the flow $f$ and each sub-flow $i_k, k\in conv(i,f,n)$, i.e., $\mathbb{P}_f(n)\cap ( \bigcup\limits_{k\in conv(i,f,n)} \mathbb{P}_{i_k})$. From this, Eq. (\ref{label28}) becomes:

\begin{eqnarray}
\label{label29}
&&  D_f^{f.ft\oplus (n-1)}(t_{f.ft\oplus (n-1)}) - D_f^{f.ft}(t_{f.ft\ominus 1})  \\
& &  \geq \min_{l \in \mathbb{P}_f(n)} (R^l - \sum \limits_{i \ni l, i \neq f} \rho_i). \nonumber \\
& & [t_{f.ft\oplus (n-1)} - t_{f.ft\ominus 1} - \sum \limits_{ l \in \mathbb{P}_f(n)}T^l \nonumber\\
&&  - \sum \limits_{i \in \mathbb{K}_f(n)} \frac{\sum\limits_{k\in conv(i,f,n)}{\sigma_i^{k\ominus 1}}+ \rho_i. \sum\limits_{j \in \mathbb{P}_f(n)\cap \mathbb{P}_i} T^{j}}{\min \limits_{l \in \mathbb{P}_f(n)} (R^l - \sum \limits_{i \ni l, i \neq f} \rho_i)} ] ^+\nonumber
\end{eqnarray}

This finishes the proof of the theorem.

\end{proof}

\newpage

%###################################
\begin{table}[h]
\caption{Notations}
\label{tab1}
\begin{center}
\begin{tabularx}{\linewidth} {l X}
\hline
    $M$   & Number of nodes in the network    \\
    $I$ & Set of flows served within the network \\
    $i\oplus k$ & $k^{th}$ node downstream from node $i$\\
    $i\ominus k$ & $k^{th}$ node upstream from node $i$ \\
    $i \ni k$ & Flow $i$ crossing the node $k$ \\
    $\mathbb{P}_i(n)$ & Subpath of flow $i$ from its source through $n$ hops, $n\leq h_i$\\
    $conv(i,f,n)$ & the convergence points of the \textit{f.o.i} $f$ with the interfering flow $i$ along its subpath of length $n$\\
        $h_i$ & Number of hops within $\mathbb{P}_i$ \\
    $\mathbb{K}_f(n)$ & Set of interfering flows with flow $f$ along $\mathbb{P}_f(n)$ \\
     $\overline{\mathbb{K}_f(n)}$ & Transformed  $\mathbb{K}_f(n)$ when cutting virtually the cycles \\
    $Mft(i, f, n)$ & First multiplexing node label of flows $i$ and $f$ along $\mathbb{P}_f(n)$ \\
    $Mlt(i,f, n)$ & Last multiplexing node label of flows $i$ and $f$ along $\mathbb{P}_f(n)$ \\
    $\beta^k(t)$ & Service curve guaranteed within node $k$ \\
     $\alpha_i^0(t)$ & Input arrival curve of flow $i$ at its initial source\\
     $\alpha_i^{k\ominus 1} (t)$ & Input arrival curve of flow $i$ at node $k$ along its path \\
     $A_i^k$ & Cumulative Arrival Function (CAF) for the flow $i$ at the node $k$\\
     $D_i^k$ & Cumulative Departure Function (CDF) for the flow $i$ at the node $k$\\
      $R^k$   & Service rate of node $k$  \\
   $T^k$   & Service latency of node $k$  \\
   $D^j$ & is the delay within the node $j$\\
   $\sigma_i^{k\ominus 1}$ & Maximum input burst of flow $i$ at node $k$ \\
	 $\rho_i$ & Maximum rate of flow i \\
	 $NP$ & Maximum priority levels, where $0$ denotes the highest one\\
	 $PL(i)$ & Priority level of flow $i$\\
	 $L_{max}(i)$ & Maximum packet length of flow $i$, accounting the communication protocol overhead\\
	 $hp_f^k$ & The set of flows crossing the node $k$ excluding the \textit{f.o.i} $f$, with priority equal or higher to the $f$ one\\
	 $lp_f^k$ & The set of flows crossing the node $k$ with priority equal or lower to the $f$ one\\
	 $\mathbb{K}_{\leq f}(n)$ & The set of flows interfering with the \textit{f.o.i} $f$ along its subpath, $\mathbb{P}_f(n)$, with a priority equal or higher to $f$ one\\
\hline
\end{tabularx}
\end{center}
\end{table}
%=========================================================================================

\newpage

\bibliographystyle{unsrt}
\bibliography{Biblio}   % name your BibTeX data base

\end{document}